\documentclass[openacc]{rsproca_new}

\usepackage{longtable}
\usepackage{bm}
\usepackage{overpic}
\usepackage{tikz}
\newcommand{\beq}{\begin{equation}}
\newcommand{\eeq}{\end{equation}}
\newcommand{\lb}{\left(}
\newcommand{\rb}{\right)}

\usepackage{url}
\usepackage{mathtools}
\usepackage{hyperref}

\titlehead{Research}

\begin{document}

\title{Anisotropy distorts the spreading of a fixed volume porous gravity current}

\author{Graham P. Benham$^{1}$}

\address{$^{1}$Mathematical Institute, University of Oxford, Oxford OX2 6GG, UK}

\subject{applied mathematics, fluid mechanics, geoscience, hydrology}

\keywords{gravity currents, porous media, anisotropy}

\corres{Graham P. Benham\\
\email{benham@maths.ox.ac.uk}}

\begin{abstract}
We consider the release and subsequent gravity-driven spreading of a finite volume of fluid in an anisotropic porous medium bounded by an impermeable substrate. When the permeability in the vertical direction is much smaller than the horizontal direction, as is the case in many real geological reservoirs, this restricts the spread of the current to a very thin layer near the impermeable base. Using a combination of asymptotic analysis and finite difference computations of Darcy flow, we show that there exist two distinct flow regimes. At early times the bulk of the current descends slowly and uniformly, injecting fluid into thin finger-like regions near the base. At much later times the current transitions to the classical gravity-driven solution and continues to spread with a self-similar shape. One interesting consequence is that the swept volume of the current grows differently depending on the anisotropy of the medium. This has important consequences for managing contaminant spills, where it is important to minimise the contacted volume of the aquifer, or during geological CO$_2$ sequestration where a larger contacted volume results in more CO$_2$ being stored. 
\end{abstract}

%\rsbreak

\begin{fmtext}
\end{fmtext} 
\maketitle

\section{Introduction}

Gravity-driven flows resulting from the release of a fluid within a porous medium are a common feature of environmental fluid dynamics. For example, such flows arise when groundwater responds to heavy rainfall \cite{guerin2014response}, after the spillage of a contaminant \cite{bear2010modeling}, or during the geological storage of carbon dioxide in saline aquifers \cite{huppert2014fluid}. Since all geological aquifers are heterogeneous and often this heterogeneity manifests as an anisotropic permeability field \cite{woods2015flow}, it is important to quantify how this affects the migration speeds and shape of the current. This is particularly relevant to situations where it is desirable to minimise the volume of the aquifer contacted by the fluid (e.g. containing the spread of a contaminant) or to maximise the contacted volume (e.g. trapping residual saturation during CO$_2$ storage).

In the case of homogeneous and isotropic porous media, there have been numerous studies on the evolution of fixed volume gravity currents. Some of these studies have treated single-phase flows \cite{huppert1995gravity}, whilst others have incorporated multiphase effects such as residual trapping \cite{hesse2008gravity,golding2017two} and dissolution within the ambient fluid \cite{macminn2011co}. In some cases simple scaling laws were derived by exploiting the self-similar properties of gravity currents, as shown by \cite{barenblatt1952some} in the case of single-phase flow, and by \cite{kochina1983groundwater} in the case of trapped residual saturation. Later work by \cite{hesse2007gravity,ball2017relaxation,zheng2023radial} explored the transition to self-similarity in both confined and unconfined settings. However, less attention has been paid to the case of heterogeneous or anisotropic porous media, despite the relevance to real geological reservoirs.

Nevertheless, some progress has been made for specific types of buoyancy driven flows in heterogeneous media. For example, several studies have investigated how anisotropy affects convective dissolution within an ambient fluid phase \cite{cheng2012effect,green2014steady,de2017dissolution}. In the case of gravity currents resulting from constant injection, \cite{jackson2020small} explored how heterogeneities of different lateral and vertical scales affect the migration speed of a CO$_2$ plume. Likewise, \cite{benham2022axisymmetric} addressed the case of a gravity current resulting from point source injection in an anisotropic medium. In this study, it was shown that anisotropy can cause a build-up of pressure that stretches the flow into an ellipsoid shape during an early-time regime of the flow, before transitioning to a gravity-dominated regime at much later times. 
However, no studies have addressed how heterogeneity affects the spreading of a released volume of fluid (i.e. in the absence of injection), despite the relevance to post-injection scenarios during CO$_2$ storage, and to post-leakage scenarios in the context of contaminant spills.

CO$_2$ storage in geological reservoirs is one of the key proposed technologies to reduce emissions and limit the effects of global warming \cite{huppert2014fluid}. In such scenarios, buoyant CO$_2$ is injected into a brine-filled  reservoir beneath an impermeable cap rock. Once the injection is switched off, the CO$_2$ rises and spreads out beneath the cap rock, with a fraction of its mass being lost to residual trapping (via the drainage/imbibition cycle) and dissolution within the surrounding brine \cite{macminn2011co,krevor2015capillary}. Hence, to quantify the trapping potential of different geological reservoirs (e.g. when choosing potential storage sites) it is important to understand how the anisotropy of the aquifer may affect the historical migration of the current across the pore space. In the context of a contaminant spillage, the objective is  to contain and minimise the spread of a harmful fluid within an aquifer. Therefore, in a similar manner to the CO$_2$ storage problem, it is necessary to quantify how and where the contaminant fluid will spread in response to the heterogeneity of the aquifer, once the leak has been closed off. 

In this study we demonstrate that anisotropy restricts the flow of the gravity current to thin finger-like regions spreading near the impermeable boundary, qualitatively similar to those predicted by other studies \cite{jackson2020small}. Due to this flow distortion, the swept volume of the current is reduced for anisotropic aquifers. This indicates that isotropic aquifers may have better potential for certain forms of CO$_2$ trapping that depend on the contacted volume of pore space. By contrast, in the case of a contaminant spill, anisotropic aquifers may help contain the spread of the fluid by restricting the flow to a reduced fraction of the porous medium.

\section{Finite release in two-dimensional anisotropic media}

\subsection{Release and subsequent dynamics}
\label{sec_early}

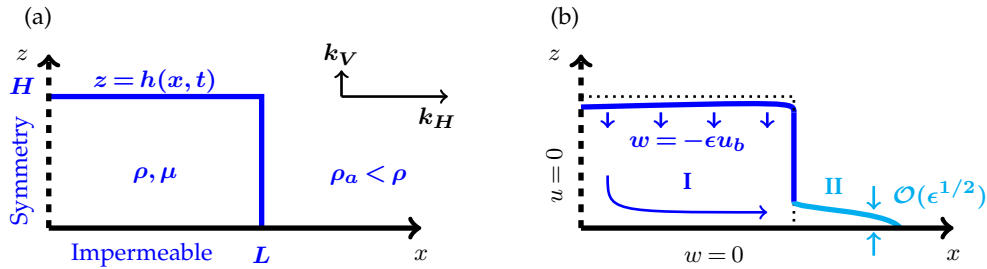
\begin{figure}
\centering
\begin{tikzpicture}[scale=0.7]
\draw [line width=2,blue] (0,2.5) -- (4,2.5) -- (4,0);
\draw[line width=2,blue] (10,2.3) .. controls (14,2.4) .. (14,2.2);
\draw[line width=2,black,->,dashed] (0,0) -- (0,3.3);
\draw[line width=2,black,->] (0,0) -- (7,0);
\draw [line width=1,dotted,black] (10,2.5) -- (14,2.5) -- (14,0);
\draw[line width=2,blue] (14,2.2) -- (14,0.5);
\draw[line width=2,cyan] (14,0.5) .. controls (14,0.35) and (15.75,0.35) .. (16,0);
\draw[line width=2,black,->,dashed] (10,0) -- (10,3.3);
\draw[line width=2,black,->] (10,0) -- (17,0);
\node at (7,-0.5) {$x$};
\node at (-0.5,3.3) {$z$};
\node[blue] at (-0.5,2.7) {$\boldsymbol{H}$};
\node[blue] at (-0.5,1.1) {\rotatebox{90}{Symmetry}};
\node[blue] at (1.75,-0.5) {Impermeable};
\node[blue] at (4,-0.5) {$\boldsymbol{L}$};
\node at (17,-0.5) {$x$};
\node at (9.5,3.3) {$z$};
\node[blue] at (2,1) {$\boldsymbol{\rho,\mu}$};
\node at (12.5,-0.5) {$w=0$};
\node at (9.5,1.) {\rotatebox{90}{$u=0$}};
\node[blue] at (12,1.65) {$\boldsymbol{w=-\epsilon u_b}$};
\draw[line width=1,blue,->] (10.5,2.2) -- (10.5,1.9);
\draw[line width=1,blue,->] (11.5,2.2) -- (11.5,1.9);
\draw[line width=1,blue,->] (12.5,2.2) -- (12.5,1.9);
\draw[line width=1,blue,->] (13.5,2.2) -- (13.5,1.9);
\node[blue] at (6,1) {$\boldsymbol{\rho_a<\rho}$};
\node[black] at (7.3,2.1) {$\boldsymbol{k_H}$};
\node[black] at (5.5,3.3) {$\boldsymbol{k_V}$};
\draw[line width=1,black,->] (5.5,2.5) -- (7.5,2.5);
\draw[line width=1,black,->] (5.5,2.5) -- (5.5,3.);
\draw[line width=1,blue,->] (10.5,1) .. controls (10.5,0.3) .. (13.5,0.3);
\node at (-0.2,4.) {(a)};
\node at (9.7,4.) {(b)};
\node[blue] at (2.,2.8) {$\boldsymbol{z=h(x,t)}$};
\node[blue] at (12.,0.9) {\bf I};
\node[cyan] at (14.75,0.8) {\bf II};
\draw[line width=1,cyan,->] (15.5,0.8) -- (15.5,0.35) ;
\draw[line width=1,cyan,->] (15.5,-0.525) -- (15.5,-0.075) ;
\node[cyan] at (16.75,0.65) {\small $\boldsymbol{\mathcal{O}(\epsilon^{1/2})}$};
%\node[blue] at (16.5,2.) {$\boldsymbol{u_b=\frac{k_H\Delta \rho g}{\mu}}$};
\end{tikzpicture}
\caption{Schematic diagram. (a) A finite volume $2LH\phi$ of dense fluid is released into a porous medium with an anisotropic permeability field $k_H\gg k_V$ above an impermeable base. (b)  The medium is initially saturated with a less dense fluid so that the bulk region (I) descends slowly at speed $-\epsilon u_b$, where $\epsilon=k_V/k_H$ and $u_b$ is the buoyancy velocity \eqref{buoy}, whilst a thin finger region (II) develops near the base of the current. \label{schem}}
\end{figure}

We consider an anisotropic porous medium in which the horizontal permeability $k_H$ is much larger than the vertical permeability $k_V$. Such anisotropic flow properties are a common feature in geological reservoirs and may result from the deposition of successive layers of fine and coarse material or from post-depositional compaction of the formation \cite{corbett1992variation}. As we will show, the anisotropy of the medium restricts the vertical flow of the bulk of the fluid, which we denote region I (see figure \ref{schem}), resulting in a slow migration towards the impermeable boundary. Meanwhile,  gravity-driven spreading is limited to thin finger-like regions near the base, which we denote region II .

To start with we restrict our attention to two-dimensional flows (although radially symmetric flows will be addressed later in Section \ref{sec_rad}) and we consider the release of a volume (per unit width) of fluid with constant density $\rho$. The surrounding porous medium is initially saturated with an ambient fluid with relatively smaller density $\rho_a<\rho$.
Due to the Boussinesq approximation \cite{soltanian2016critical}, these results also apply in the case of a lighter fluid (e.g. CO$_2$) released within a porous medium saturated with a heavier fluid (e.g. brine), with the impermeable boundary located above rather than below.

For the sake of simplicity we take the viscosity of the released fluid and the ambient fluid to be the same ($\mu$), and we consider that the initial shape of the released fluid is rectangular, with dimensions $2L\times H$. It should be noted, however, that these results would apply to any similar convex shape, as shown in Figure \ref{num_app} and discussed in more detail in Appendix \ref{sec_appa}. 

The flow in the released volume of fluid is subject to the two-dimensional Darcy equations,
\begin{align}
\mathbf{u}&=-\frac{1}{\mu}\underline{\underline{\mathbf{k}}}\cdot \nabla \left[ p + \rho g z \right],\label{dar1}\\
\nabla \cdot \mathbf{u} &=0\label{dar2},
\end{align}
where $\mathbf{u}$ is the Darcy velocity vector, $p$ is the pressure, and $\underline{\underline{\mathbf{k}}}=\mathrm{diag}(k_H,k_V)$ is the anisotropic permeability field in the $x,z$ directions. Combining \eqref{dar1} and \eqref{dar2}, the pressure satisfies Laplace's equation with anisotropic coefficients, 
\beq
\frac{\partial^2 p}{\partial x^2}+\epsilon\frac{\partial^2 p}{\partial z^2}=0,\label{lapl}
\eeq
where the anisotropy is given by
\beq
\epsilon=\frac{k_V}{k_H}\ll 1.
\eeq
The boundary conditions for the flow in region I are as follows. The left hand and bottom boundaries, $x=0$, $z=0$, are assumed to be symmetric and impermeable, respectively, so we prescribe no normal flow,
\begin{align}
u=0:& \quad x=0,\label{side}\\
w=0:& \quad z=0.\label{bottom}
\end{align}
Likewise, at the fluid interface, which we denote $z=h(x,t)$, we impose the dynamic and kinematic boundary conditions:
\begin{align}
p=p_a-\rho_a g h: & \quad z=h(x,t),\label{dyn}\\
w=\phi\frac{\partial h}{\partial t}+u\frac{\partial h}{\partial x}:& \quad z=h(x,t).\label{kin}
\end{align}
The former condition matches the pressure in the fluid with the ambient hydrostatic pressure (note that the reference pressure $p_a$ is the ambient value at $z=0$), whilst the latter condition imposes that a particle at the interface remains at the interface \cite{gilmore2022leakage,benham2022near}.

In the limit $\epsilon\rightarrow 0$, mass conservation \eqref{dar2} indicates that if there is no vertical flow (i.e. $k_V=0$) then there cannot be any horizontal flow either. Essentially, the governing equation \eqref{lapl} implies that the pressure (which is continuous) is set by the ambient fluid, such that $p=p_a-\rho_a g z$, and the resulting velocities are $u=w=0$. Hence, the interface remains at the initial position $z=h_0(x)$ for all time.

Now let's consider the case of a strongly anisotropic porous medium, such that $0<\epsilon\ll 1$. 
%In this case, the pressure is still hydrostatic (set by the ambient fluid) at leading order.
In this case, the solution can be found by performing an asymptotic expansion of the pressure in powers of $\epsilon$. 
Within this expansion, the leading order contribution to the pressure is simply the solution to the isotropic problem (i.e. with $\epsilon=0$ exactly). Hence, for the same reasons as described above, the leading order pressure is hydrostatic and set by the ambient fluid, $p=p_a-\rho_a g z$. 
However, by inserting this pressure into Darcy's law \eqref{dar1}, we now derive a small but finite (i.e. first order) vertical velocity within region I, such that
\beq
w=- \epsilon u_b,\label{wvel}
\eeq 
where 
\beq
u_b=\frac{k_H\Delta\rho g}{\mu}, \label{buoy}
\eeq
is the buoyancy velocity and $\Delta \rho = \rho-\rho_a$. 
Since \eqref{wvel} does not satisfy the impermeability condition \eqref{bottom}, it is necessary to reevaluate the solution near $z\approx 0$ using boundary layer theory.

Mathematically speaking, \eqref{lapl} is a singular perturbation problem since it is a second order Partial Differential Equation (PDE) with a small parameter in front of a second derivative. This indicates that not all vertical boundary 
conditions can be satisfied by the leading order solution. Specifically, the dynamic boundary condition \eqref{dyn} is imposed at $z=h$ to ensure continuity of pressure, whilst the impermeability condition \eqref{bottom} at $z=0$ is left unsatisfied. To correct this requires rescaling the solution to investigate changes over a small vertical distance near $z\approx 0$, also known as a boundary layer. By inspection of \eqref{lapl} it is clear that $z$ must be rescaled by a factor of $\epsilon^{1/2}$ to recover all terms in the governing equation. Hence, an appropriate choice of rescaled dimensionless variables is
\beq
x=L\xi ,\quad z= \epsilon^{1/2}L\zeta, \quad p = p_a-\rho_a g z +\epsilon^{1/2} \Delta \rho g L  \,P(\xi,\zeta),\label{rescalvars}
\eeq
where $\xi$, $\zeta$ and $P$ are variables which are $\mathcal{O}(1)$ in magnitude. 
Note that the pressure in \eqref{rescalvars} must also be rescaled by a factor $\epsilon^{1/2}$ so that the boundary condition \eqref{bottom} (i.e. $\partial p/\partial z+\rho g =0$ at $z=0$) balances all terms at leading order. 
In this way, within the boundary layer region the governing equations and boundary conditions \eqref{lapl},\eqref{side} and \eqref{bottom} become
\begin{align}
\frac{\partial^2 P}{\partial \xi ^2}+\frac{\partial^2 P}{\partial \zeta^2}&=0,\label{lapl2}\\
\frac{\partial P}{\partial \xi}&=0:\quad \xi=0,\\
\frac{\partial P}{\partial \zeta}&=-1:\quad \zeta=0.
\end{align}
In addition, we require that the \textit{inner} solution (within the boundary layer) matches with the \textit{outer} solution (far outside the boundary layer), such that
\beq
P\rightarrow 0:\quad \zeta\rightarrow \infty.\label{liminf}
\eeq
The system is not yet complete since the governing equation \eqref{lapl2} is a second order elliptic PDE which requires four boundary conditions. The fourth and final boundary condition needs more careful thought. Since there is a vertical velocity \eqref{wvel} descending through the outer region, this induces an arrival of flux $\epsilon u_b L$ within the inner region. However, since this flux can go neither downwards nor leftwards (due to impermeable/symmetric boundaries), it must instead exit through the right hand boundary, $x=L$. In other words,  the right hand interface must move outwards to conserve mass, creating a new finger-like region of vertical size $z\sim \mathcal{O}(\epsilon^{1/2})$, which we denote region II.
Hence, the final boundary condition for region I is given by an integral constraint of the form
\beq
\int_0^\infty -\left.\frac{\partial P}{\partial \xi}\right|_{\xi=1} \, \mathrm{d}\zeta =1.\label{flux}
\eeq
The descent of the upper interface in region I and the resulting finger-like region II are both illustrated in figure  \ref{schem}b.

Before addressing these details further, we first note that \eqref{lapl2}-\eqref{flux} can be solved exactly by separation of variables. Hence, the composite solution (valid across both inner and outer regions) is given by
\beq
p = p_a-\rho_a g z + 2\epsilon^{1/2}\Delta \rho g L \sum_{n=0}^\infty \frac{(-1)^n }{\lambda_n^2} \cos \left[\frac{\lambda_n x}{L}\right] \exp\left[-\frac{\lambda_n z}{\epsilon^{1/2}L}\right],\label{fullsol}
\eeq
where $\lambda_n=(2n+1)\pi/2$.
Conservation of mass within region I indicates that
\beq
\phi \frac{\partial h}{\partial t} + \frac{k_H}{\mu} \frac{\partial }{\partial x}\left[ \int_0^{h(x,t)} -\frac{\partial p}{\partial x} \, \mathrm{d}z \right]=0.\label{masscon}
\eeq
Hence, inserting \eqref{fullsol} into \eqref{masscon} results in the governing equation for the evolution of the thickness of the current, which is
\beq
\phi \frac{\partial h}{\partial t} -   u_b \epsilon L\frac{\partial }{\partial x}  \sum_{n=0}^\infty \frac{2(-1)^{n+1}\sin\lambda_n x/L}{\lambda_n^2}   \left( 1-\exp \left[-\frac{\lambda_n h(x,t)}{\epsilon^{1/2} L}\right]\right) =0.\label{masscon2}
\eeq
This can be further simplified by ignoring exponentially small terms when $h/L$ is larger than $\mathcal{O}(\epsilon^{1/2})$, and by using the fact that the infinite sum converges to $-x/L$.
Hence, we see that the thickness within region I is given by
\beq
h=H-   \frac{\epsilon u_b t}{\phi} ,\label{lintime}
\eeq
which is valid for $0<x<L$ and for $h\gg\mathcal{O}(\epsilon^{1/2})$.

\begin{figure}
\centering
\begin{tikzpicture}[scale=1.1]
\node at (0,0) {\includegraphics[width=0.32\textwidth]{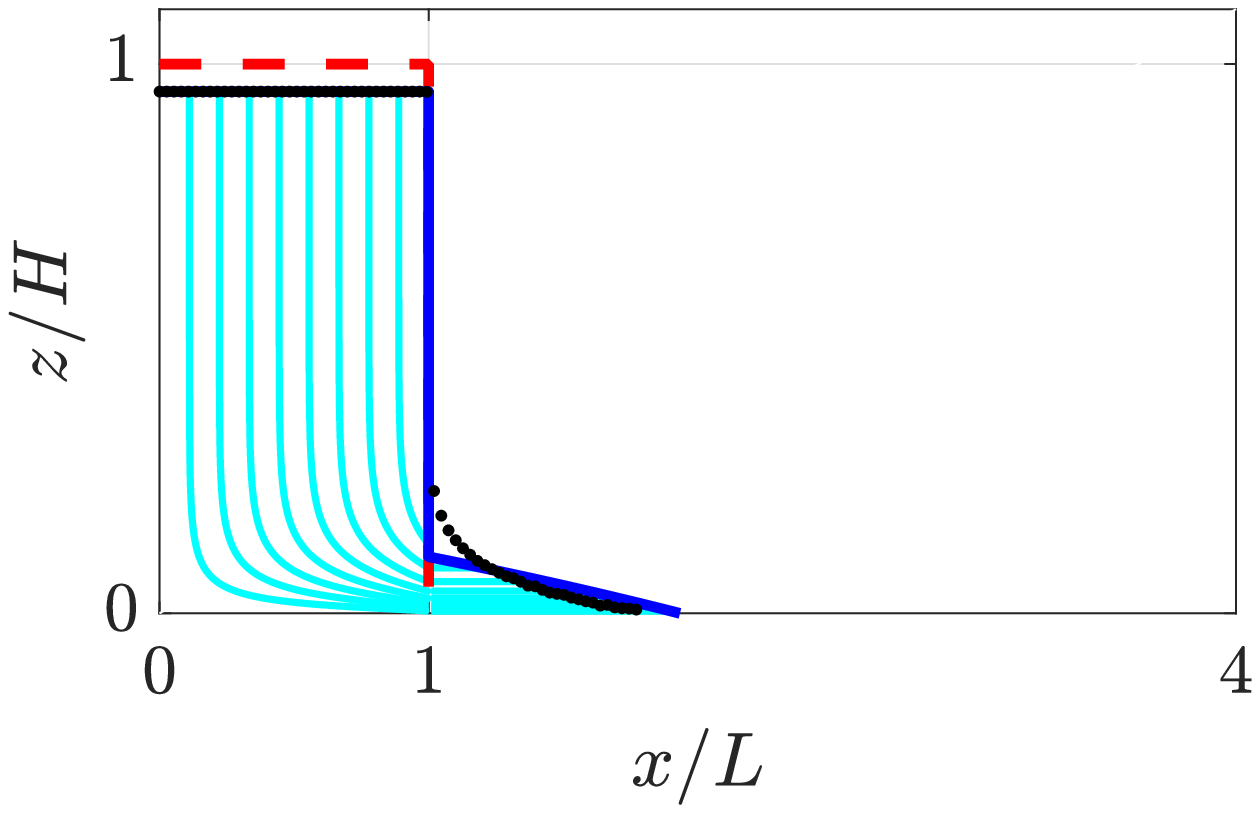}};
\node at (4.1,0) {\includegraphics[width=0.32\textwidth]{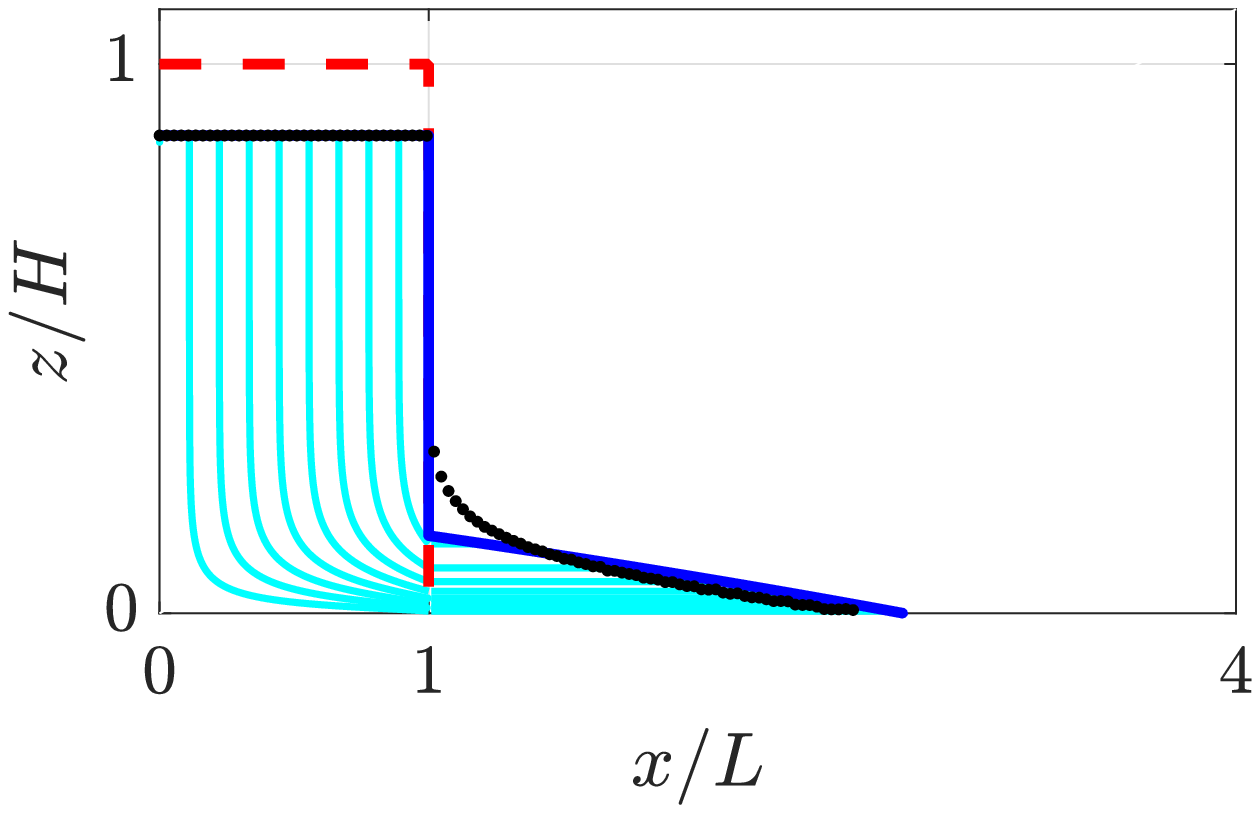}};
\node at (8.25,0) {\includegraphics[width=0.32\textwidth]{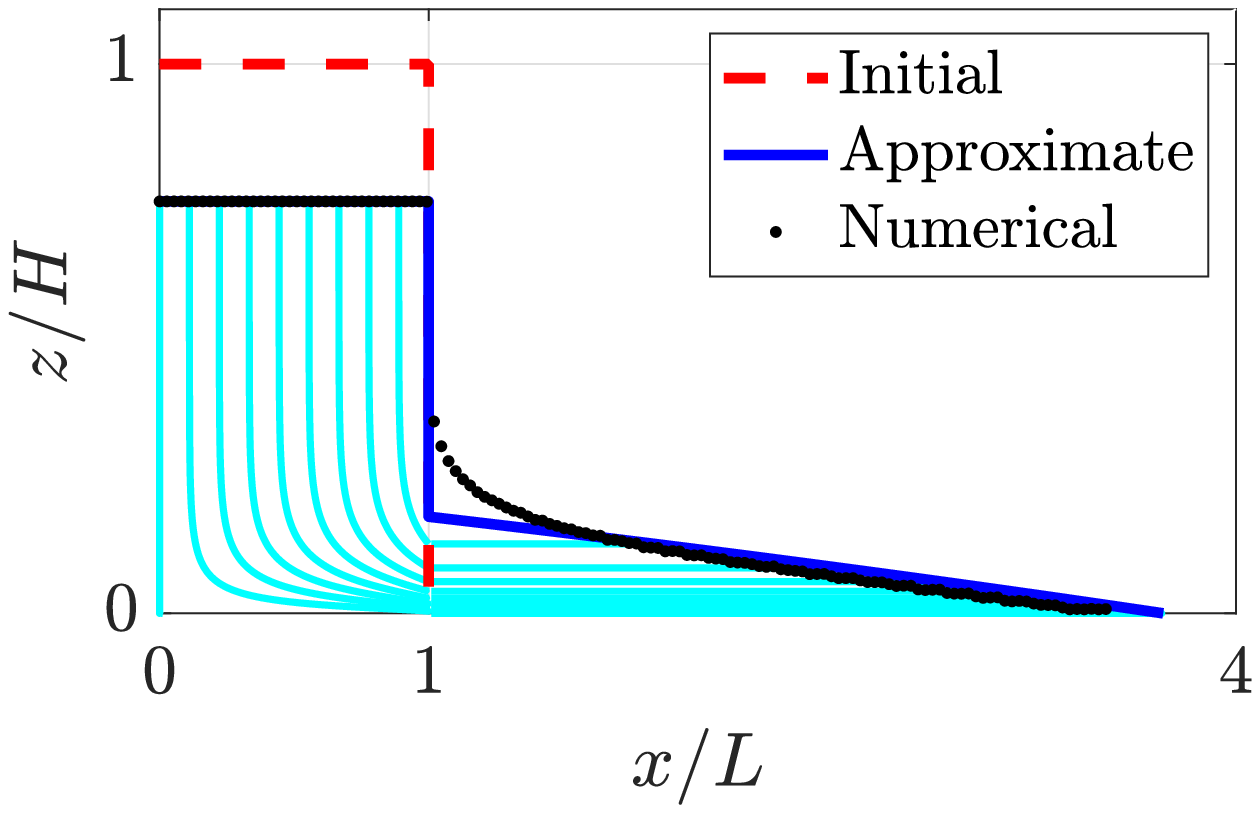}};
\node at (-0.7,-2.2) {\includegraphics[width=0.15\textwidth]{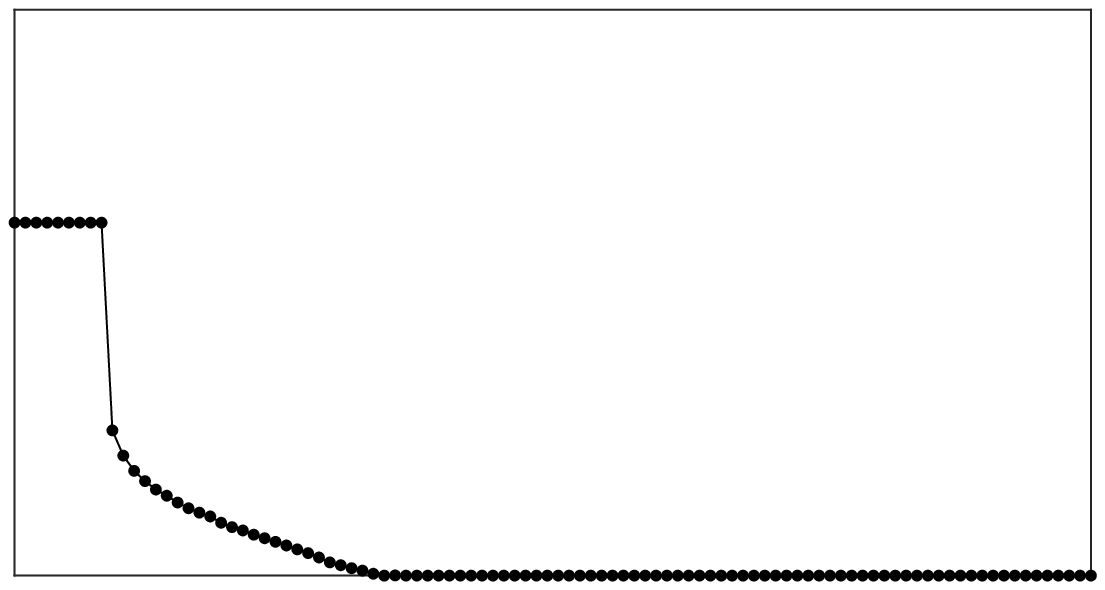}};
\node at (1.8,-2.2) {\includegraphics[width=0.15\textwidth]{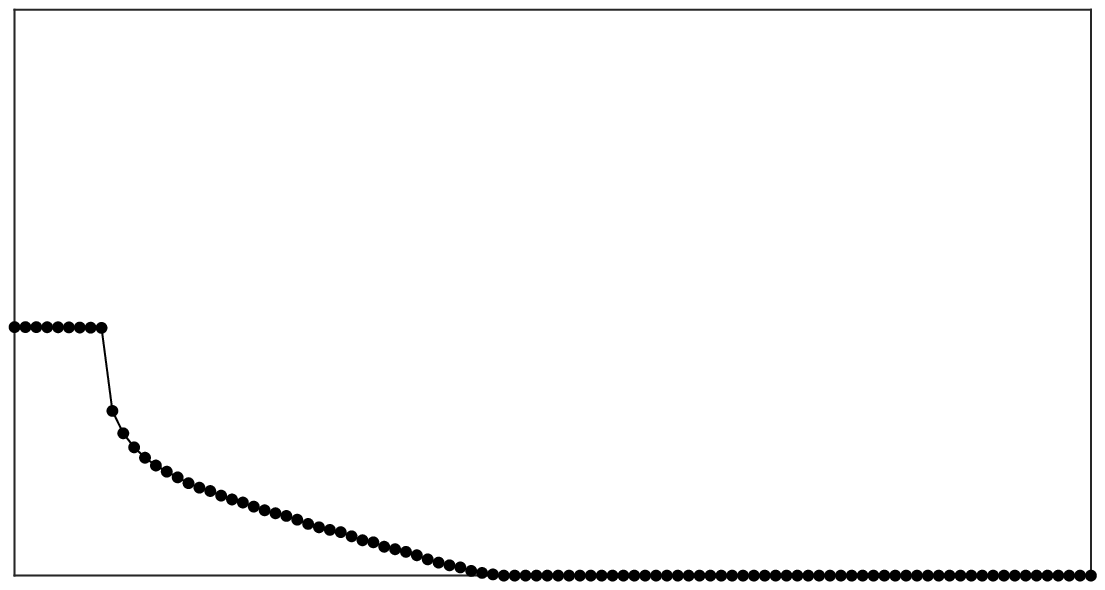}};
\node at (4.3,-2.2) {\includegraphics[width=0.15\textwidth]{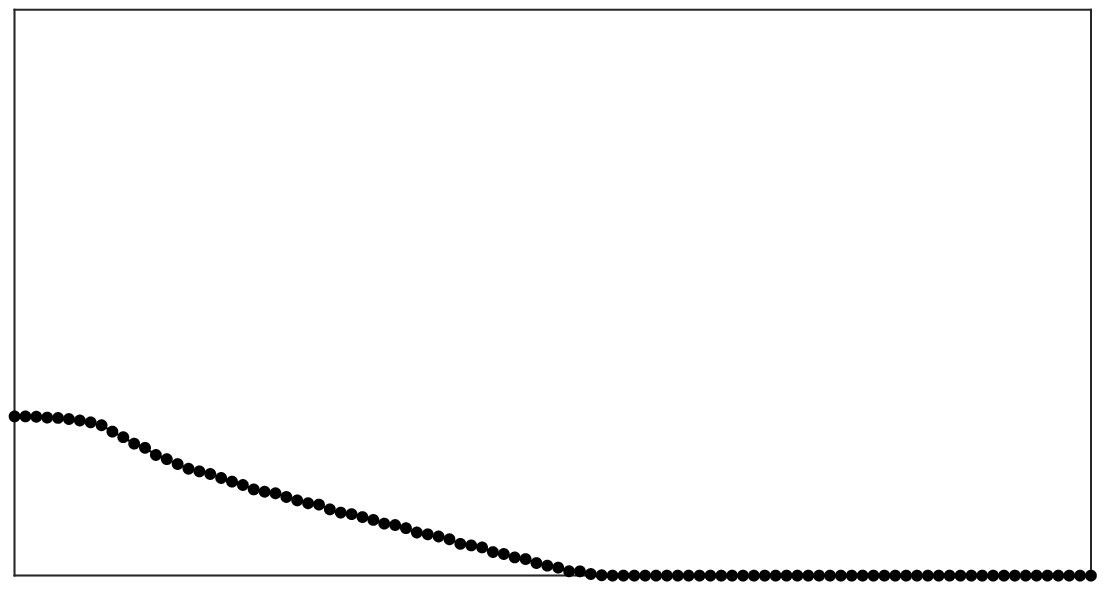}};
\node at (6.8,-2.2) {\includegraphics[width=0.15\textwidth]{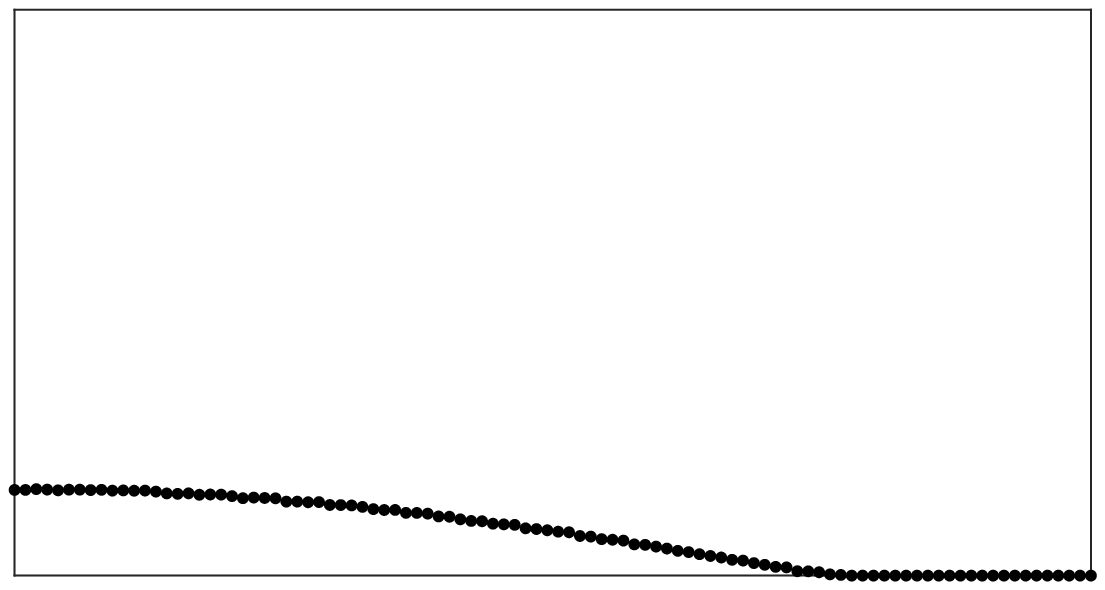}};
\node at (9.3,-2.2) {\includegraphics[width=0.15\textwidth]{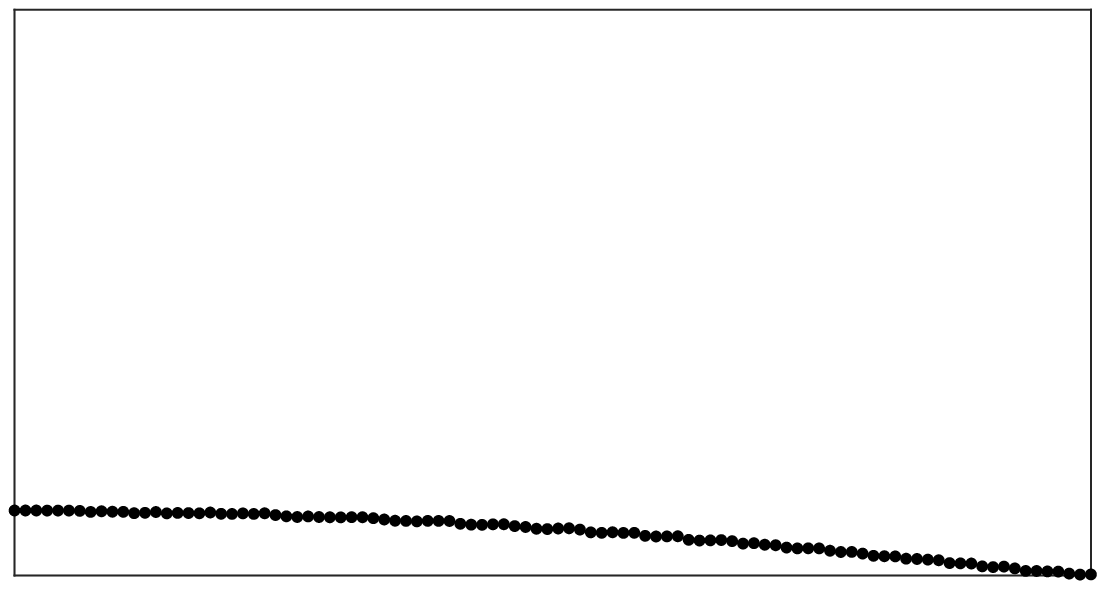}};
\node at (-1.6,1.6) {(a)};
\node at (2.7,1.6) {(b)};
\node at (6.8,1.6) {(c)};
\node at (-1.8,-1.4) {(d)};
\node[red] at (-0.4,0.5) {\bf I};
\node[red] at (0,-0.4) {\bf II};
\end{tikzpicture}
\caption{(a,b,c) Approximate solution for regions I and II compared with numerical solution in the case of $\epsilon=10^{-2}$. Times are given in terms of the transition time \eqref{transt}, where (a) $t/t^*=0.07$, (b) $0.17$ and (c) $0.34$. 
Streamlines are evaluated using velocities derived from \eqref{fullsol} and \eqref{darcyII}. (d) Further numerical solutions are shown at times $t/t^*=0.42, 0.69, 1, 1.82, 3.22$. 
\label{num_an}}
\end{figure}

Next, we address the fluid flow in region II, which is the finger-like region of escaped fluid near the base of the current, which is defined for $L<x<x_N(t)$, where $x_N(t)$ is the maximum extent of the finger. This flow region is long and thin (like a classical gravity current) such that the horizontal velocity is much larger than the vertical velocity. Consequently, the pressure within region II is hydrostatic to good approximation, such that
\beq
p=p_a-\rho_a g h - \rho g\lb z-h \rb.\label{darcyII}
\eeq
Similarly to \eqref{masscon}, conservation of mass within this region gives
\beq
\phi \frac{\partial h}{\partial t} =u_b \frac{\partial }{\partial x}\left[ h \frac{\partial h}{\partial x}  \right],\label{massconIII}
\eeq
which is sometimes called the Dupuit approximation. This is accompanied by boundary conditions that correspond with imposing the input flux from region I, 
\beq
-u_b  h \frac{\partial h}{\partial x}= \epsilon u_b L:\quad x=L, \eeq
and imposing zero thickness and zero flux at the moving front,
\begin{align}
h\rightarrow  0:\quad x\rightarrow x_N(t),\label{hbc1}\\
-u_b  h \frac{\partial h}{\partial x} \rightarrow 0:\quad x\rightarrow x_N(t).\label{hbc2}
\end{align}
By introducing dimensionless coordinates, 
\beq
h=\epsilon^{1/2} L\mathcal{H},\quad x= L(1+\xi),\quad t=\frac{\phi  L}{ \epsilon^{1/2} u_b} \tau,\label{simvars1}
\eeq
we get the same system as \cite{huppert1995gravity} for a constant input flux (see Appendix \ref{sec_appb} for further details).
The solution is well known and is given in terms of the similarity variables
\beq
\eta=\xi/\tau^{2/3},\quad \mathcal{H}=\tau^{1/3}f(\eta).\label{earlysim}
\eeq
The self-similar shape function $f(\eta)$ is defined for $\eta\in[0,\eta_N=1.482]$, and is monotone decreasing from $f_0:=f(0)=1.296$ to $f(\eta_N)=0$. The solution for regions I and II is plotted in figure \ref{num_an}a,b,c, at several different times. Streamlines confirm that the flux into region II is fed by the shrinking of region I. Comparison is also made to a numerical solution described later in Section \ref{sec_num}.

\begin{figure}
\centering
\begin{tikzpicture}[scale=1.1]
\node at (1.5,-4.5) {\includegraphics[width=0.5\textwidth]{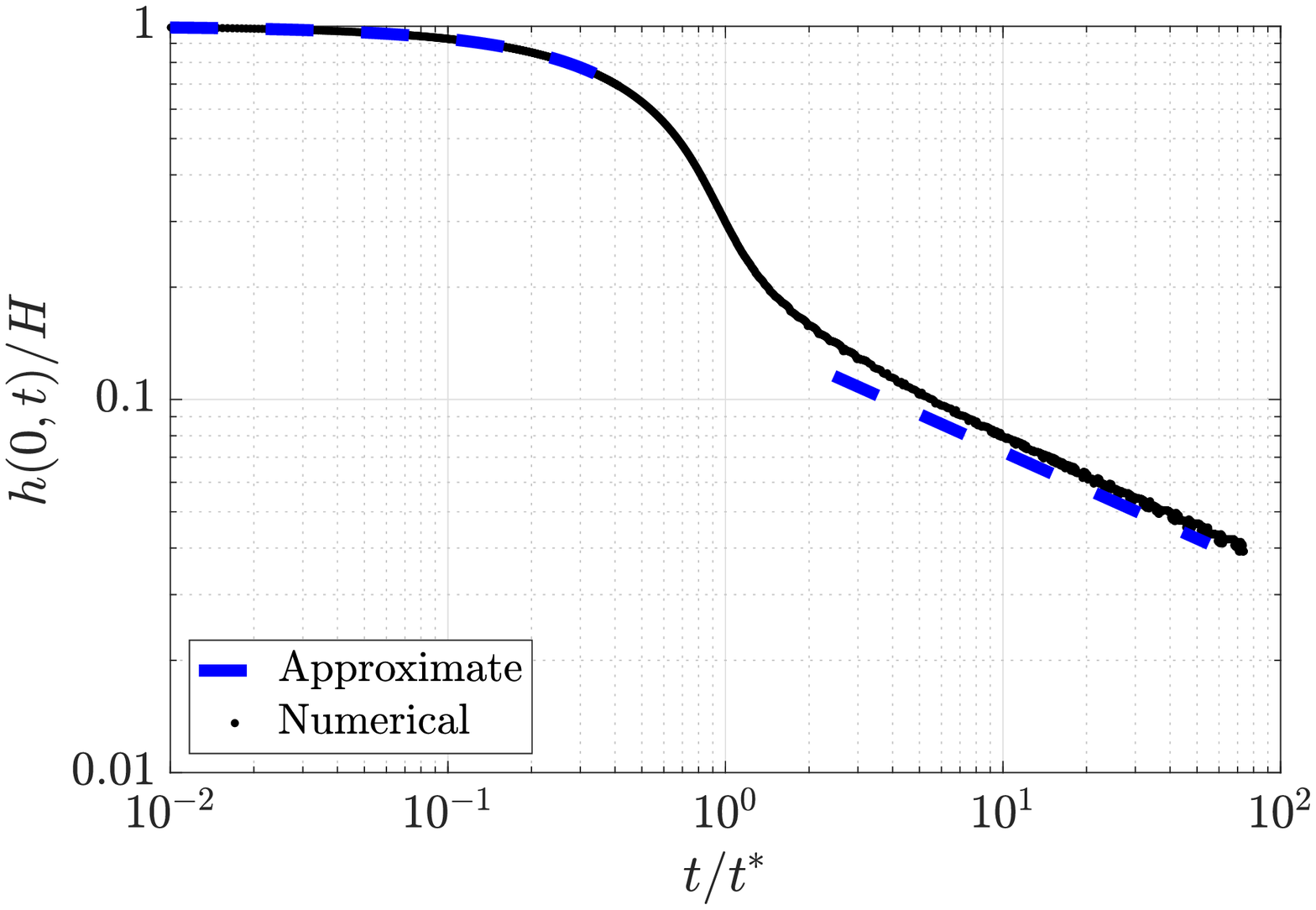}};
\node at (7.7,-4.5) {\includegraphics[width=0.5\textwidth]{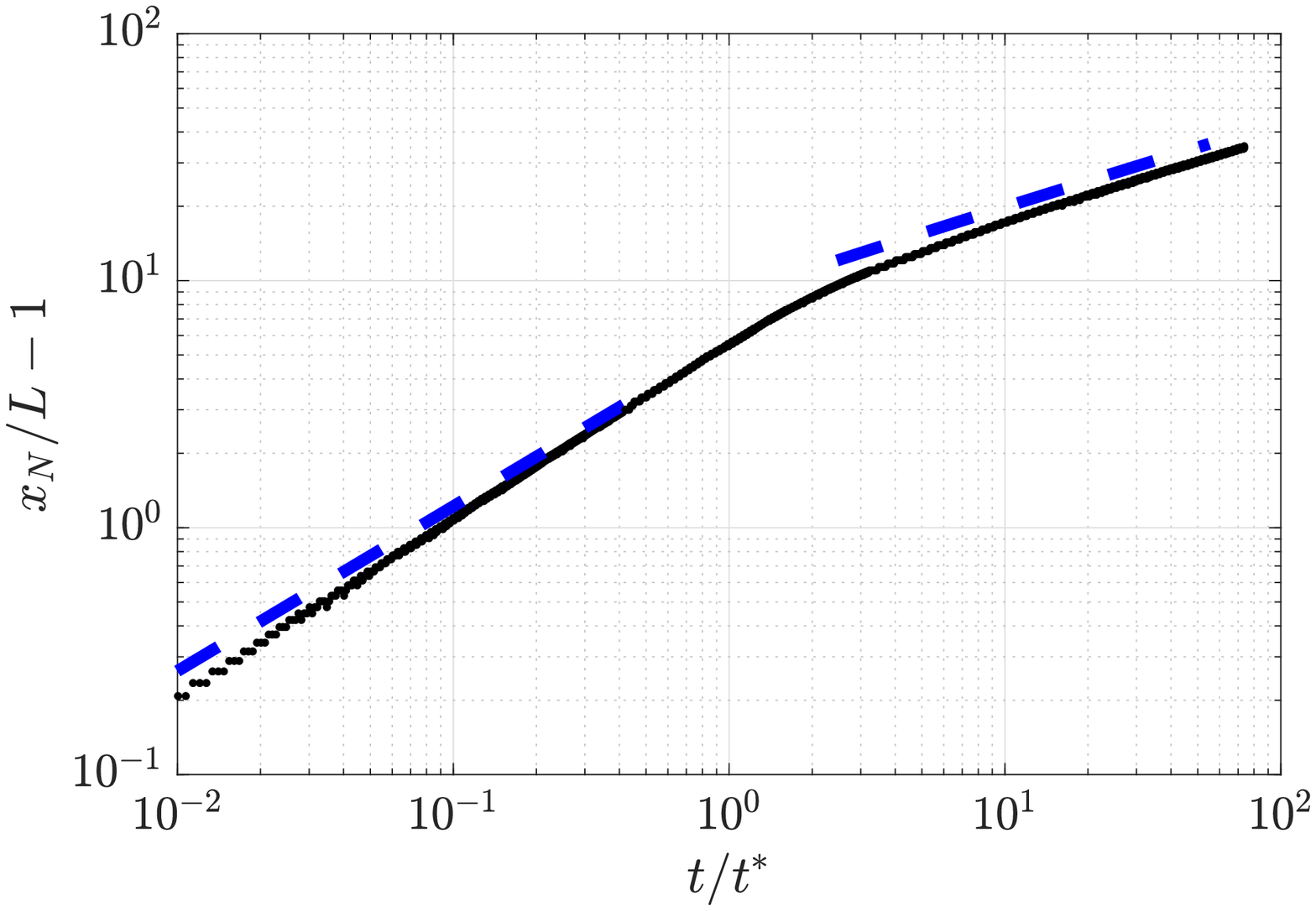}};
\node at (-1.,-2.) {(a)};
\node at (5.1,-2.) {(b)};
\end{tikzpicture}
\caption{ Evolution of the vertical (a) and horizontal (b) extents of the gravity current shown in figure \ref{num_an} (i.e. for $\epsilon=10^{-2}$), showing both numerical and approximate solutions. Early time behaviour is given by \eqref{lintime},\eqref{xNearly}, whereas late time behaviour is given by \eqref{Hlate},\eqref{xNlate}. 
%Approximate behaviour is given by $h(0)/H=1-(\epsilon u_b t/H\phi)$ and $\xi_N/\eta_N\alpha^{2/3}=(\epsilon u_b t/H\phi)^{2/3}$.
\label{num_2}}
\end{figure}

The maximum vertical extent of the flow is given by \eqref{lintime}, whereas the maximum horizontal extent is determined through the above scalings as
\beq
x_N(t)=L(1+  \eta_N \tau^{2/3}). \label{xNearly}
\eeq
These are plotted in figure \ref{num_2} with dashed blue lines, thereby indicating the early-time behaviour of the released fluid. 
%As before, there is good agreement with the finite difference computations. 

\subsection{Transition to self-similarity}\label{sec_trans}

After a long time the flow is expected to eventually transition to the self-similar behaviour of a finite release gravity current  \cite{huppert1995gravity,ball2017relaxation}. As discussed by \cite{benham2022axisymmetric}, the late-time dynamics of a two-dimensional gravity current are independent of the anisotropy of the medium since the bulk flow decouples from the ambient. Hence, the anisotropy only affects the late-time behaviour by delaying the time to transition to self-similarity. Hence, to study the late-time behaviour we first analyse the isotropic case, which determines the late-time dynamics, and then use these dynamics to derive the time $t^*$ to transition between the early and late solutions, where $t^*$ depends on the anisotropy $\epsilon$.

As before, the thickness of the gravity current (which now occupies a single region $0\leq x\leq x_N(t)$) satisfies the Dupuit approximation \eqref{massconIII}.
The boundary conditions are similar to the case of constant input flux, as described above, except the left hand boundary condition is replaced with the zero flux condition
\beq
\frac{\partial h}{\partial x}= 0:\quad x=0.\label{symbc}
\eeq
Consequently, mass conservation indicates that
\beq
\int_0^{x_N(t)} h \, \mathrm{d}x=HL.\label{intconst}
\eeq
By introducing dimensionless coordinates, 
\beq
x=(HL)^{1/2}\xi,\quad h=(HL)^{1/2}\mathcal{H},\quad t=\frac{\phi (HL)^{1/2}}{u_b}\tau, \label{finitescle0}
\eeq
and switching to similarity variables
\beq
\eta =\xi/\tau^{1/3},\quad \mathcal{H}=f(\eta)/\tau^{1/3},\label{finitescle}
\eeq
we arrive at a system of equations that can be solved analytically to give
\beq
f(\eta)=\frac{1}{6}(\eta_N^2-\eta^2),\label{analf}
\eeq
where $\eta_N=3^{2/3}$, as shown by \cite{huppert1995gravity} (see Appendix \ref{sec_appb} for further details). 
%, and this is plotted in \cite{huppert1995gravity}. 

The maximum vertical and horizontal extent of the flow are given by 
\begin{align}
  h(0,t)&=\frac{\eta_N^2(HL)^{1/2}}{6\tau^{1/3}},\label{Hlate}\\
x_N(t)&=\eta_N(HL)^{1/2}\tau^{1/3}\label{xNlate},
\end{align} 
respectively. These scalings are plotted in figure \ref{num_2} with dashed blue lines, thereby indicating the late-time behaviour of the gravity current. 
%As before, good agreement is obtained when compared with numerical simulations. 

Next, we discuss the time taken to transition from the early flow regime involving two fluid regions to the late flow regime with a single region which is self-similar. In the early regime the vertical extent of the flow descends according to \eqref{lintime}. By contrast, the vertical extent of region II ($\epsilon^{1/2} L \mathcal{H}$ in \eqref{simvars1}) increases like $\sim f_0 (u_b \epsilon^2 L^2 t/\phi)^{1/3}$. Hence, it is expected that the transition to self-similarity will occur once these two thickness scalings approach each other. Thus, the transition time $t^*$ satisfies the cubic equation
\beq
\left(H- \frac{\epsilon u_b t^*}{\phi}\right)^3=\frac{ f_0^3 u_b \epsilon^2 L^2 t^*}{\phi}.\label{transsolve}
\eeq
The solution has a complicated form but can be expanded in powers of $\epsilon\ll1$ to give
\beq
t^*=\frac{L\phi}{u_b\epsilon }\left[\alpha-f_0 \alpha^{1/3} \epsilon^{1/3} +\frac{f_0^2}{3\alpha^{1/3}} \epsilon^{2/3} + \ldots \right],\label{transt}
\eeq
where $\alpha=H/L$ is the aspect ratio of the initial current shape. 
Hence, anisotropy delays the transition to a classical self-similar regime, which is consistent with other studies \cite{benham2022axisymmetric}. At the transition time $t^*$ the maximum thickness of the current (which we denote $H^*$) is given by
\beq
H^*=L\epsilon^{1/3}\left[f_0 \alpha^{1/3}  -\frac{f_0^2}{3\alpha^{1/3}} \epsilon^{1/3} + \ldots \right].\label{transh}
\eeq
This indicates that, by the time transition occurs for very anisotropic media, the bulk of the current has shrunk significantly. This hints towards a reduced swept volume, which we analyse further in Section \ref{sec_sweep}.

\begin{figure}
\centering
\begin{tikzpicture}[scale=1.1]
\node at (0,0) {\includegraphics[width=0.5\textwidth]{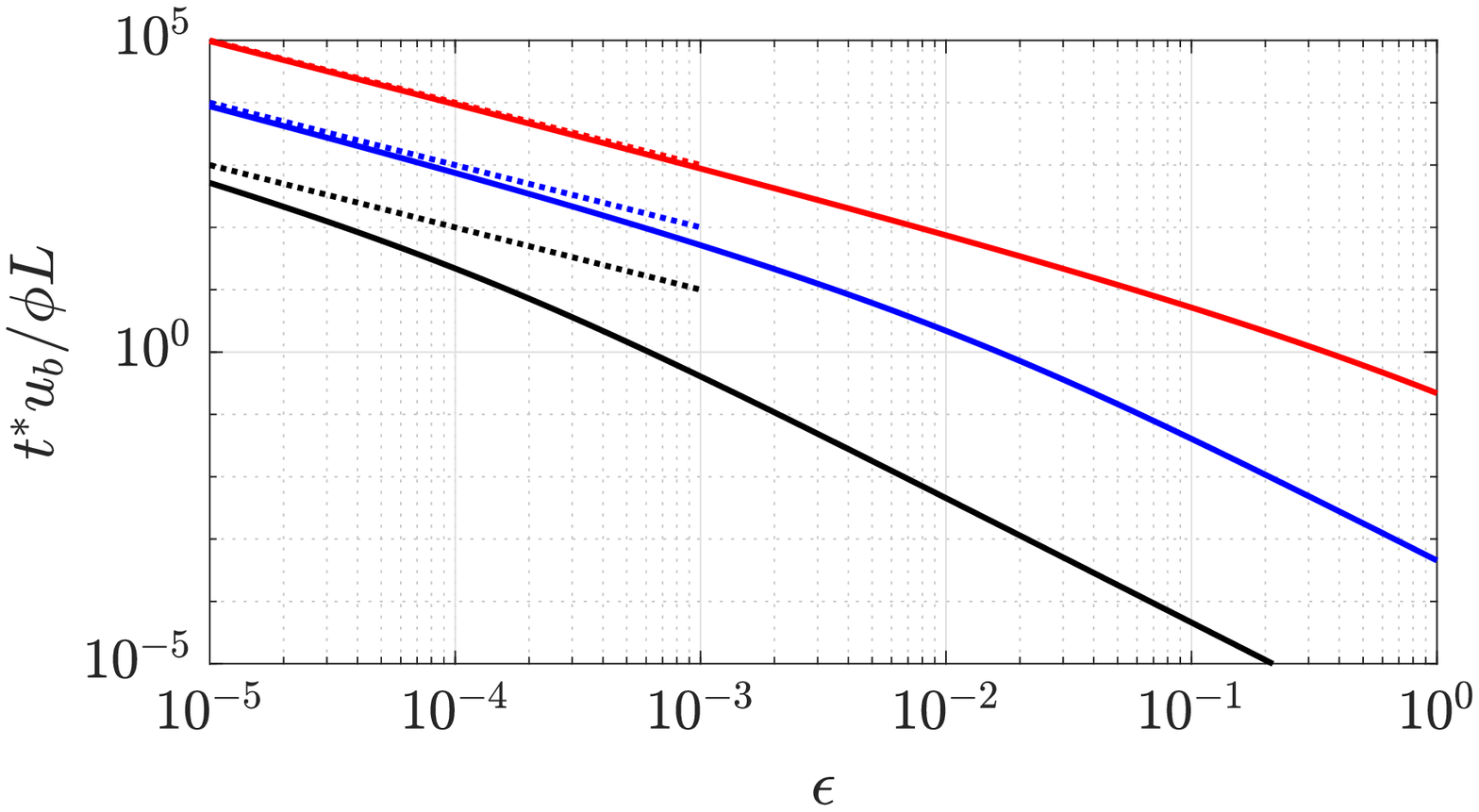}};
\node at (6.2,0) {\includegraphics[width=0.5\textwidth]{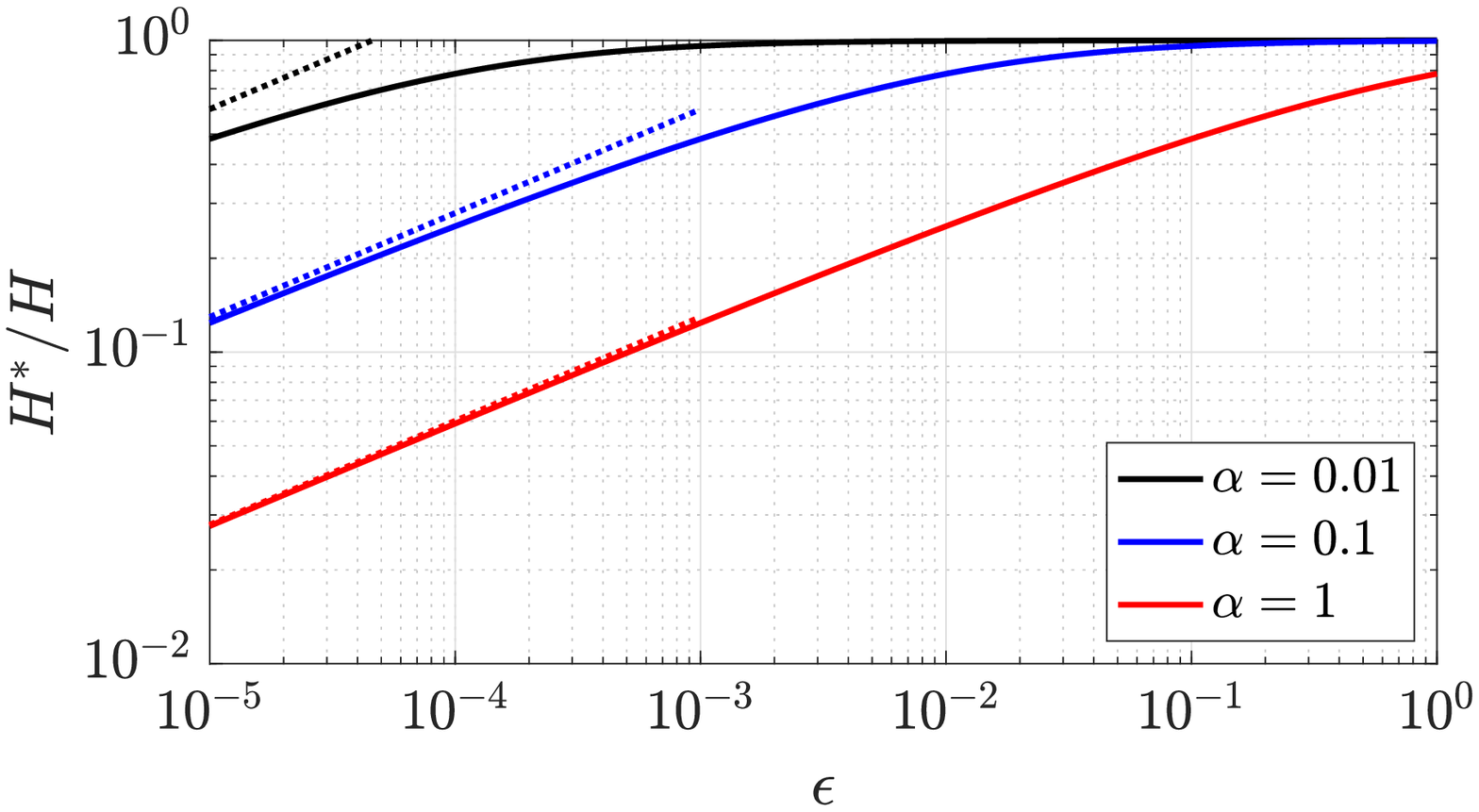}};
\node at (-2.2,2.) {(a)};
\node at (3.5,2.) {(b)};
\end{tikzpicture}
\caption{Transition time $t^*$ and the transition thickness $H^*$, determined by solving \eqref{transsolve} for different values of the anisotropy $\epsilon$ and the initial aspect ratio $\alpha$. Leading order scalings for small $\epsilon$ (see \eqref{transt},\eqref{transh}) are shown with dotted lines. \label{transdata}}
\end{figure}

The transition time $t^*$ and the transition thickness $H^*$ (given in dimensionless terms) are plotted in figure \ref{transdata}. These importantly depend on both the anisotropy $\epsilon$ as well as the initial aspect ratio of the flow $\alpha$. It should be noted that in practice strong anisotropy $\epsilon\ll1$ may cause a stretching of the flow before contact with the impermeable cap rock. Therefore, $\epsilon\ll1$ is likely to be correlated with $\alpha\ll1$.

\section{Finite difference computations of Darcy flow}\label{sec_num}

Next, in this section we compare our analytical predictions to finite difference computations of two-dimensional Darcy flow. The flow is modelled with the Darcy equations \eqref{dar1}-\eqref{dar2}, accounting for different permeability values, $k_H$ and $k_V$, in the horizontal and vertical directions.
The governing equations are accompanied by boundary conditions \eqref{side}-\eqref{bottom}, corresponding with impermeable/symmetric walls at $x=0$ and $z=0$. The dynamic boundary condition \eqref{dyn} is imposed on the interface $z=h(x,t)$, which is interpolated over a gridded mesh of 150$\times$150 points. Likewise, a spatial domain of finite size $4L\times H $ is chosen in the $x\times z$ directions. The fluid flow is not resolved beyond the interface $z>h(x,t)$ since pressure is assumed to be hydrostatic in the ambient fluid. Mass conservation dictates that the interface evolves according to 
\beq
\phi \frac{\partial h}{\partial t} + \frac{k_H}{\mu} \frac{\partial }{\partial x}\left[ \int_0^{h(x,t)} -\frac{\partial p}{\partial x} \, \mathrm{d}z \right]=0,\label{massconnum}
\eeq
with suitable initial conditions $h(x,0)=h_0(x)$.
Boundary conditions for $h$ are given by \eqref{hbc1}, \eqref{hbc2} and \eqref{symbc} (see earlier discussion for further explanation). 
The time-dependent equation for the fluid-fluid interface \eqref{massconnum} is solved using an explicit forward Euler scheme in time, and a backward eighth order scheme in space. At each time-step the Darcy equations, \eqref{dar1}-\eqref{dar2}, are solved using a second order central finite difference scheme.
The code used for these computations is available in the Supplementary Materials. 

Results from the finite difference computations are compared with the analytical model in figures \ref{num_an} and \ref{num_2} for an anisotropy value $\epsilon=0.01$. Figure \ref{num_an}a,b,c shows the distorted spreading of the gravity current via a thin finger near the base. Good agreement is observed everywhere except near $x\approx L$ where the two regions connect. Here, the interface transitions smoothly between regions I and II, which is a second order feature that is missing from the simple analytical model. The numerical solution enables computation of the gravity current shape up to and beyond the transition time $t^*$, as is displayed in figure \ref{num_an}d. This demonstrates clearly how the flow transitions from a two-region (bulk/finger) structure at early times to a slumping single-region structure at late times. 

Figure \ref{num_2} displays numerical computations of the vertical and horizontal extents of the gravity current across early and late time regimes. Clearly, the numerical computations reflect the transition between the early and late time analytical scalings, shown with dashed lines. Likewise, the transition between these different regimes corresponds with $t/t^*\approx 1$, indicating the accuracy of our prediction for $t^*$ \eqref{transt}.

As a benchmark test, we also compare these finite difference computations with a gravity current slumping in an isotropic medium, $\epsilon=1$. Since our analytical model only applies for $\epsilon\ll1$, the initial dynamics involving regions I and II are not relevant for this case. Instead, we set the initial shape as
\beq
h_0(x)=H\left[1-\left(\frac{x}{L}\right)^2\right],\label{initsim}
\eeq
which is simply the similarity solution \eqref{analf}. Since the initial shape satisfies the similarity conditions, this ensures that the solution remains self-similar for all time. In figure \ref{num_app}a,b,c, in Appendix \ref{sec_appa}, we  plot the self-similar evolution of the gravity current shape at various times. Excellent agreement is attained between the numerical model and the exact self-similar solution, indicating the reliability of the finite difference approach. In figure \ref{num_app}d,e,f, we display similar computations for the same initial shape released in an anisotropic medium, $\epsilon=0.01$. 
In this case, the flow is decomposed into regions I and II, as before, demonstrating how our simple model can be extended to account for other released shapes. Further details and discussion of this case are given in Appendix \ref{sec_appa}.

\section{Swept shape and swept volume}
\label{sec_sweep}

\begin{figure}
\centering
\begin{tikzpicture}[scale=1.1]
\node at (0,0) {\includegraphics[width=0.5\textwidth]{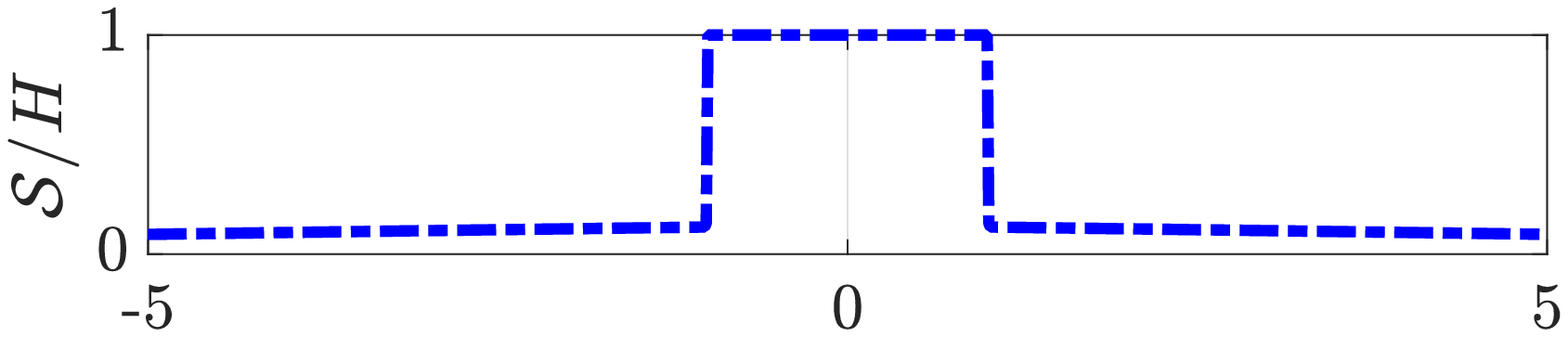}};
\node at (0,-1.5) {\includegraphics[width=0.5\textwidth]{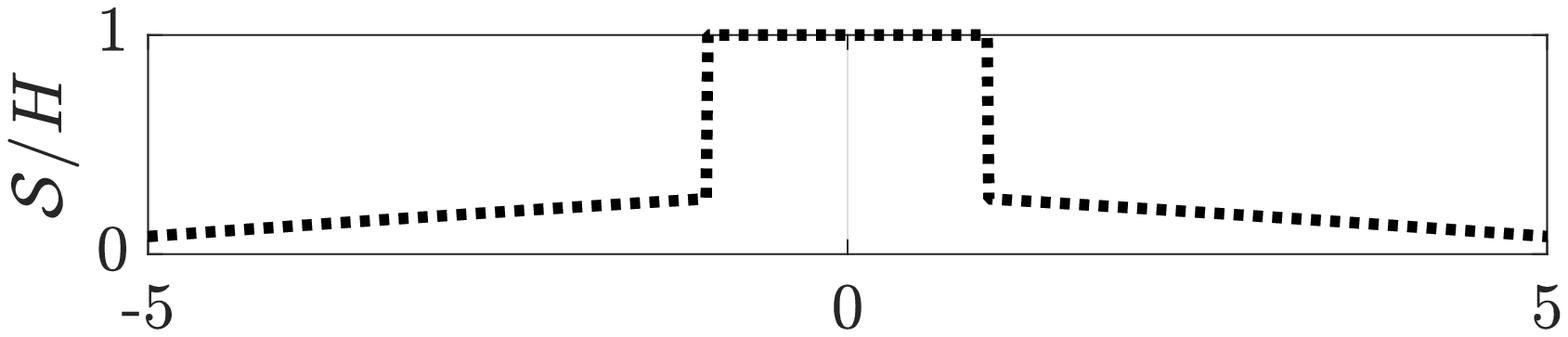}};
\node at (0,-3) {\includegraphics[width=0.5\textwidth]{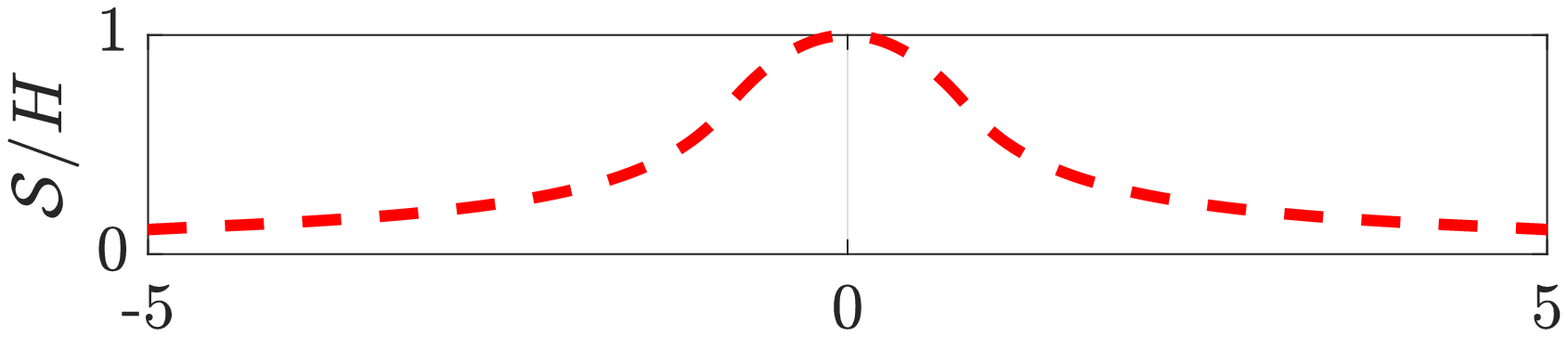}};
\node at (6.1,-1.8) {\includegraphics[width=0.45\textwidth]{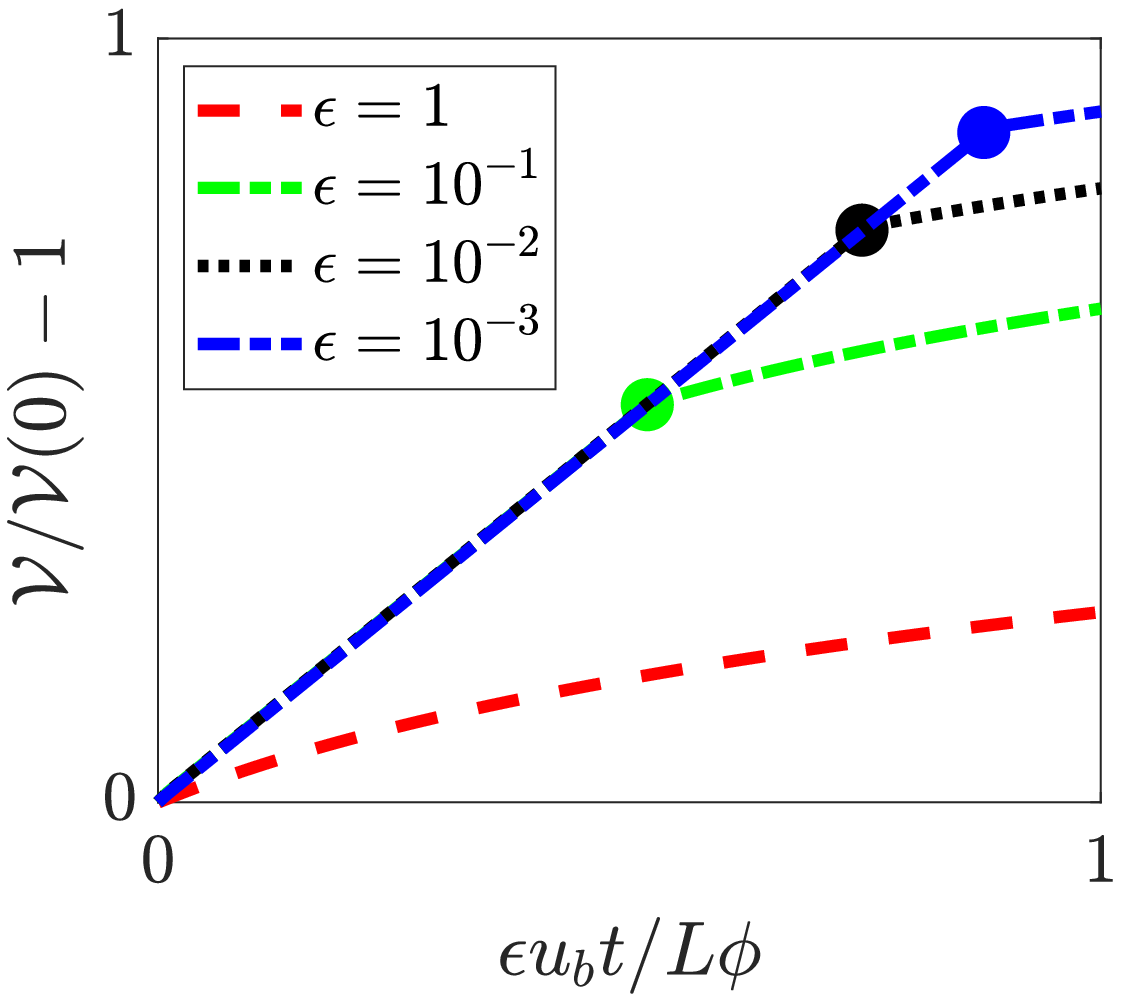}};
\node at (-3.,0.8) {(a)};
\node at (-3.,-0.8) {(b)};
\node at (-3.,-2.3) {(c)};
\node at (3.6,0.9) {(d)};
\node at (0.3,-4) {$x/L$};
\end{tikzpicture}
\caption{(a,b) Swept shape of the gravity current $\mathcal{S}(x)$ \eqref{sweepfun} in the case of $\epsilon=10^{-3}$ and $10^{-2}$. (c) Self-similar isotropic case $\epsilon=1$ (note the modified initial conditions). (d) Evolution of the swept volume $\mathcal{V}(t)$ for each of these cases (note that time is stretched by a factor $\epsilon$). Transition times $t^*$ are indicated with coloured dots. The initial aspect ratio is set as $\alpha=1$ in all cases. \label{sweep}}
\end{figure}

As described earlier, it is useful for applications (e.g. contaminant transport or CO$_2$ storage) to quantify the total volume contacted by the gravity current, also known as the swept volume\footnote{Note that since our model is in two dimensions, the swept volume is taken as per unit width. Also note that the total volume of fluid is constant and given by $2HL\phi=\phi\mathcal{V}(0)$.}. 
At early times $t\ll t^*$ region I shrinks uniformly downwards whilst region II grows upwards and outwards. Hence, the swept volume $\mathcal{V}$ is simply equal to the initial volume plus the instantaneous volume of region II, such that
\beq
\mathcal{V}(t)=2HL+2\epsilon u_b L t. \label{Aearly}
\eeq
At much later times $t\gg t^*$, once the gravity current has transitioned to self-similarity, the thickness $z=h(x,t)$ has some parts which are growing and other parts which are shrinking, so this requires more careful attention.
To deal with this we first define the swept shape $\mathcal{S}$ as the maximum thickness that the current ever reached at a given value of $x$, such that
\beq
\mathcal{S}(x)=\max_t\left\{ h(x,t) \right\}.\label{sweepfun}
\eeq
The swept volume is then given in terms of $\mathcal{S}$ as
\beq
\mathcal{V}(t)=2\int_{0}^{x_N(t)} \mathcal{S} \,\mathrm{d} x.
\eeq
The swept shape \eqref{sweepfun} is calculated by finding the time at which the thickness is maximal, which is equivalent to $\eta=3^{1/6}$ in the similarity solution $f$ \eqref{analf}. Inserting this into \eqref{sweepfun} we get
\beq
\mathcal{S}(x)= \frac{HL}{3^{1/2}x},
\eeq
which is only valid for $x\gg L$. Hence, the swept volume (up to a constant $C$) is
\beq
\mathcal{V}(t)= \frac{2HL}{3^{1/2}}\log x_N(t) + C,\label{Alate}
\eeq
which clearly diverges like $\sim\log t$ as $t\rightarrow\infty$.
The constant of integration is found by equating \eqref{Aearly} and \eqref{Alate} at the transition time $t^*$. Hence, the late time behaviour of the swept volume is given by
\beq
\mathcal{V}(t)= 2HL + 2\epsilon u_b L t^* + \frac{2HL}{3^{1/2}} \log \left(\frac{t}{t^*}\right)^{1/3}.\label{Alate2}
\eeq
In figure \ref{sweep}a,b,d, the swept shape $\mathcal{S}(x)$ and swept volume $\mathcal{V}(t)$ are plotted for different values of the anisotropy $\epsilon$. For $\epsilon\ll1$, the anisotropy restricts the swept shape (for $x>L)$ to finger-like regions near the base of the current. The maximum thickness of the fingers is set by $H^*$ \eqref{transh}, which scales like $H^*\sim \epsilon^{1/3}$ (i.e. the more anisotropic the medium, the narrower the fingers). Furthermore, anisotropy delays the transition time \eqref{transt}, since $t^*\propto \epsilon^{-1}$. Meanwhile, before transition occurs, $t\ll t^*$, anisotropy causes the swept volume to grow slowly (since $\mathcal{V}\propto \epsilon t$ in \eqref{Aearly}), such that the gravity current contacts a smaller fraction of pore space at early times. 
Note that the kinks at the corners $x=\pm L$ (in $\mathcal{S}$) and at the transition time $t=t^*$ (in $\mathcal{V}$) would be smoothed out by higher order asymptotics or numerical computations of two-dimensional Darcy flow. However, such features do not affect the overall leading order behaviour displayed here.

For comparison, we also plot the swept shape $\mathcal{S}$ in the isotropic case, $\epsilon=1$, in figure \ref{sweep}c. 
%Since our model only applies for $\epsilon\ll1$, the initial dynamics involving regions I and II are not relevant for this case. Instead, we set the initial shape as
%\beq
%h_0(x)=H\left[1-\left(\frac{x}{L}\right)^2\right],\label{initsim}
%\eeq
%which is simply the similarity solution \eqref{analf} (note the initial swept volume is then modified to $\mathcal{V}(0)=4HL/3$). Since the initial shape satisfies the similarity conditions, this ensures that the solution remains self-similar for all times. 
As described at the end of Section \ref{sec_num}, the isotropic case uses \eqref{initsim} as the initial shape and remains self-similar for all times. It should be noted that the modified initial shape results in a different initial swept volume for this case, $\mathcal{V}(0)=4HL/3$. 
%The corresponding swept shape $\mathcal{S}(x)$ is plotted with red dashed lines in figure \ref{sweep}a,b, for comparison with the anisotropic cases. 
Clearly, the classical (isotropic) self-similar solution has a larger swept shape and a faster growing swept volume than the anisotropic cases. In Section \ref{sec_disc}, we discuss how these results can be applied when modelling contaminant spills and CO$_2$ sequestration, noting the possible limitations of our model. 

\section{Finite release with radial symmetry}
\label{sec_rad}

The above results can be easily extended to account for radially symmetric flows if the anisotropy remains aligned with the vertical coordinate, i.e. with permeability $k_H,k_V$ in the radial/vertical directions. To extend the model from the two-dimensional case, we take the impermeable boundary to be the horizontal plane, $z=0$.  
Likewise, we consider the released shape to be a cylinder of initial height $H$ and radius $R$. Since the initial shape of the current is radially symmetric, it will spread out and remain radially symmetric for all times. 

Following the derivation in Section \ref{sec_early}, the thickness of the current evolves according to \eqref{lintime} at early times. This results in a radial flux of magnitude $ \epsilon u_b \pi R^2$ exiting region I ($0\leq r \leq R$) into a growing annular region II ($R\leq r\leq r_N(t)$) at the base of the current.
Conservation of mass within  region II gives
\beq
\phi \frac{\partial h}{\partial t} =u_b \frac{1}{r}\frac{\partial }{\partial r}\left[ r h \frac{\partial h}{\partial r}  \right].\label{massconrad}
\eeq
Boundary conditions correspond with imposing the input flux from region I, 
\beq
-2\pi u_b r h \frac{\partial h}{\partial r}=  \epsilon u_b \pi R^2:\quad r=R, \label{originbc}
\eeq
and imposing zero thickness and zero flux at the moving front,
\begin{align}
h\rightarrow  0:\quad r\rightarrow r_N(t),\label{hbcrad1}\\
-2\pi r u_b  h \frac{\partial h}{\partial r} \rightarrow 0:\quad r\rightarrow r_N(t).\label{hbcrad2}
\end{align}
By introducing dimensionless coordinates, 
\beq
h=(\pi\epsilon)^{1/2} R\mathcal{H},\quad r= R(1+\xi),\quad t=\frac{\phi  R}{ (\pi \epsilon)^{1/2} u_b} \tau,\label{simvarsrad1}
\eeq
we get a similar system to \cite{lyle2005axisymmetric} (see Appendix \ref{sec_appb} for further details).
The solution is given in terms of the similarity variables
\beq
\eta=\xi/\tau^{1/2},\quad \mathcal{H}=f(\eta).\label{earlysimrad}
\eeq
The shape function $f(\eta)$ is defined for $0<\eta\leq \eta_N=1.155$, but has an unphysical singularity at the origin.  This singularity, which is due to a breakdown of the hydrostatic assumption, can be addressed by introducing a non-hydrostatic (i.e. source-driven) region near the origin, as discussed in \cite{benham2022axisymmetric}. 
However, for the sake of simplicity, we ignore such details for the present study. For our purposes, \eqref{earlysimrad} serves as a good approximation for the finger-like growth of region II (i.e. for $r\gg R$). The transition time $t^*$ is found by matching the two thicknesses of regions I and II, such that
\beq
H- \frac{\epsilon u_b t^*}{\phi}=(\pi\epsilon)^{1/2}R.\label{transsolverad}
\eeq
Hence, the transition time and thickness are given by
\begin{align}
    t^*&=\frac{\phi R}{\epsilon u_b}\left(\alpha - (\pi\epsilon)^{1/2}\right),\label{tstarrad}\\
    H^*&=(\pi\epsilon)^{1/2} R,
\end{align}
where in this case the initial aspect ratio is $\alpha=H/R$. 
Much later than the transition time $t\gg t^*$, the flow continues to slump as a single region, similar to Section \ref{sec_trans}. In this case, the entire thickness satisfies \eqref{massconrad}.
The boundary condition \eqref{originbc} is replaced by a zero flux condition at the origin, which is 
\beq
\frac{\partial h}{\partial r}=0:\quad r=0.\label{symbcrad}
\eeq
Hence, mass conservation indicates that
\beq
2\pi \int_0^{r_N(t)} r h \, \mathrm{d}r=\pi H R^2.\label{intconstrad}
\eeq
By introducing dimensionless coordinates, 
\beq
r=(H R^2)^{1/3}\xi,\quad h=(H R^2)^{1/3}\mathcal{H},\quad t=\frac{\phi (H R^2)^{1/3}}{u_b}\tau, \label{finitesclerad0}
\eeq
and switching to similarity variables
\beq
\eta =\xi/\tau^{1/4},\quad \mathcal{H}=f(\eta)/\tau^{1/2},\label{finitesclerad}
\eeq
we arrive at a system of equations that can be solved analytically to give
\beq
f(\eta)=\frac{1}{8}(\eta_N^2-\eta^2),\label{analfrad}
\eeq
where $\eta_N=2$ (see Appendix \ref{sec_appb} for further details). 

\begin{figure}
\centering
\begin{tikzpicture}[scale=1.1]
\node at (0,0) {\includegraphics[width=0.45\textwidth]{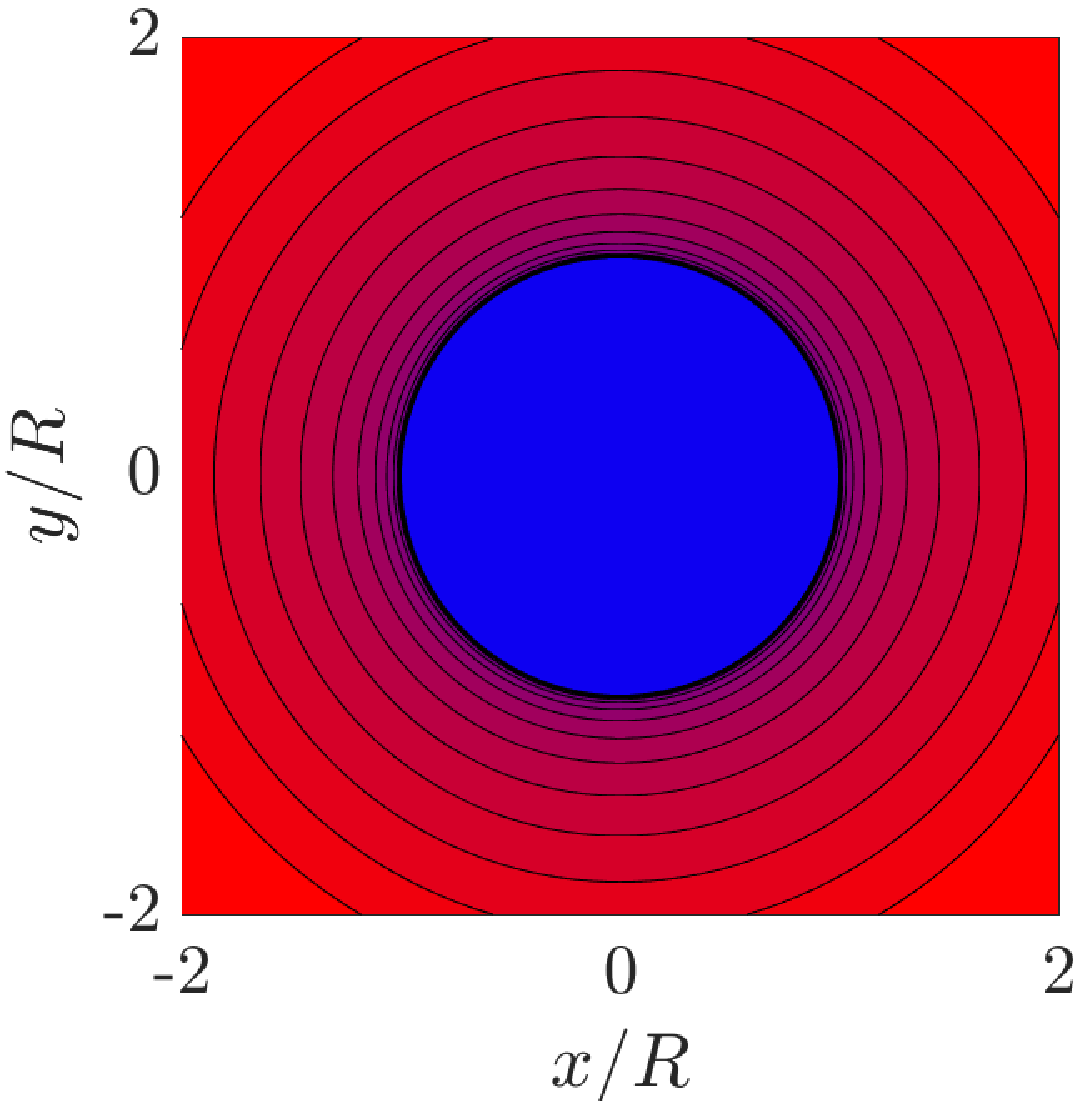}};
\node at (6.2,0) {\includegraphics[width=0.5\textwidth]{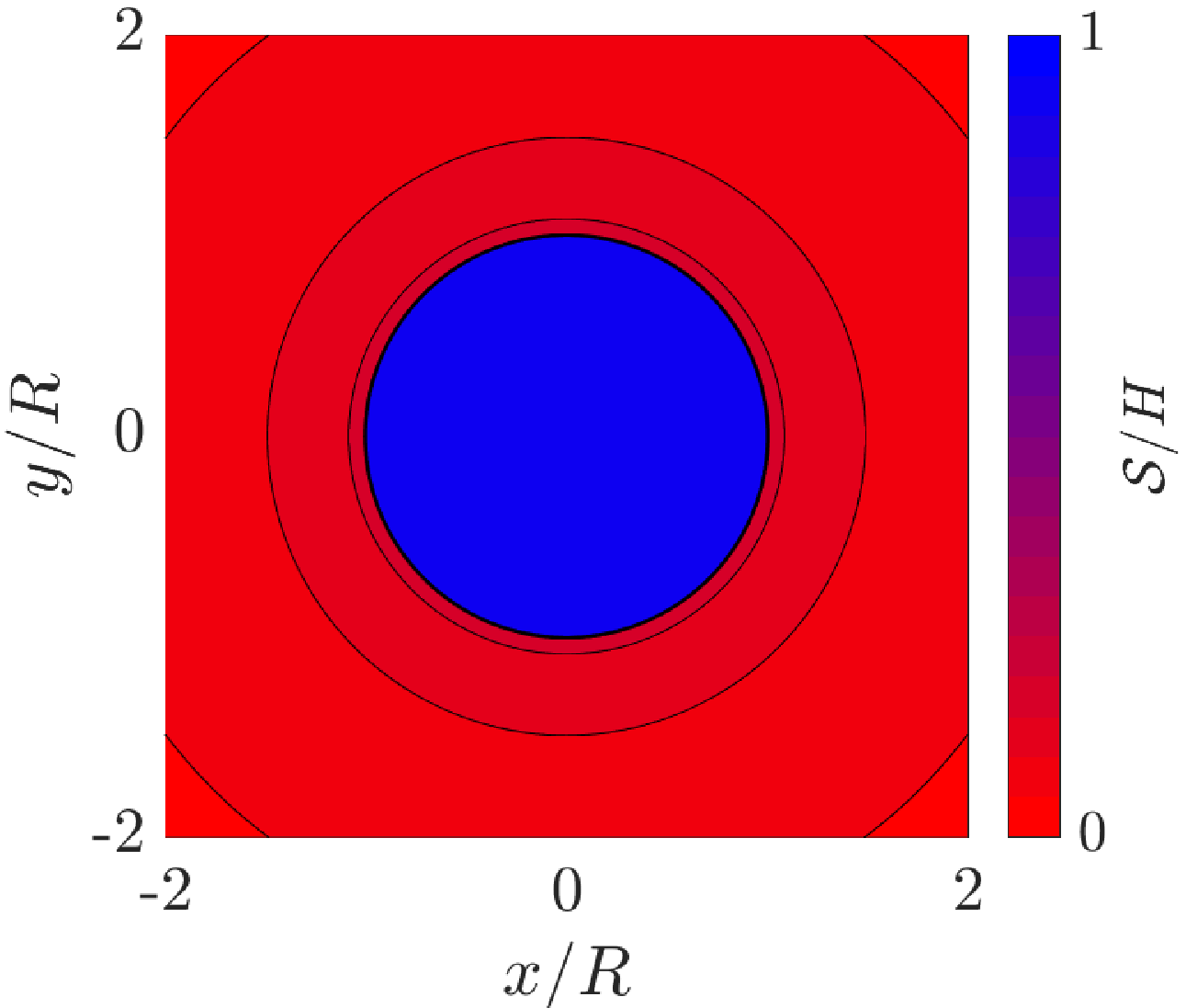}};
\node at (-2.6,2.4) {(a)};
\node at (3.35,2.4) {(b)};
\end{tikzpicture}
\caption{Swept shape of the gravity current $\mathcal{S}(r)$ in the radially symmetric case. Initial aspect ratio is set to $\alpha=0.2$ and the anisotropy is set to (a) $\epsilon=10^{-2}$ and (b) $\epsilon=10^{-3}$. \label{colorplts}}
\end{figure}

Next, let's briefly discuss the swept shape and swept volume for this case. At early times $t\ll t^*$, the swept volume $\mathcal{V}$ is equal to the initial volume plus the instantaneous volume of region II, such that
\beq
\mathcal{V}(t)=\pi H R^2+ \epsilon u_b \pi R^2 t. \label{Aearlyrad}
\eeq
At much later times $t\gg t^*$, the swept shape $\mathcal{S}(r)$ is calculated by finding the time at which the current thickness is maximal, which is equivalent to $\eta=2^{1/2}$ in the similarity solution $f$ \eqref{analfrad}. Hence, the swept shape is
\beq
\mathcal{S}(r)= \frac{H R^2}{2 r^2},
\eeq
which is only valid for $r\gg R$. The swept volume, which is now defined as
\beq
\mathcal{V}(t)=2\pi \int_{0}^{r_N(t)} r\mathcal{S} \,\mathrm{d} r.
\eeq
is calculated (following similar steps as in Section \ref{sec_sweep}) for late times as
\beq
\mathcal{V}(t)= \pi H R^2 + \epsilon u_b \pi R^2 t^* + \pi HR^2 \log \left(\frac{t}{t^*}\right)^{1/4}.\label{Alate2rad}
\eeq
To illustrate the radially symmetric case, the swept shape $\mathcal{S}(r)$ is plotted in figure \ref{colorplts}a,b, for $\epsilon=10^{-2}$ and $10^{-3}$. As with the two-dimensional case, stronger anisotropy results in a reduced swept shape (i.e. smaller values of $\mathcal{S}$ for $r>R$). Likewise, although not plotted, the swept volume increases slowly like $\mathcal{V}\propto \epsilon t $ before transition to self-similarity, indicating that anisotropy reduces the contacted volume of pore space.

\section{Discussion and applications}
\label{sec_disc}

The gravity-driven spreading of a finite volume of fluid in anisotropic porous media differs from the isotropic case since the vertical flow is restricted by the permeability $k_V\ll k_H$. In the initial dynamics the bulk of the flow descends slowly with a boundary layer near the impermeable base that diverts the flow into thin finger-like regions growing slowly in the lateral direction. This partition of the flow into bulk and finger regions reduces the swept volume of the gravity current compared to the isotropic case. This indicates that released volumes in anisotropic aquifers contact a smaller fraction of the available pore space. Hence, the spread of a contaminant in an anisotropic aquifer may be easier to contain since the contacted volume is reduced. By contrast, in the case of CO$_2$ sequestration, where the aim is to trap as much CO$_2$ in the pore space as possible, isotropic aquifers may have better potential in terms of residual trapping (which is a function of the contacted volume of pore space), though this ignores other trapping mechanisms such as dissolution, structural trapping and mineralisation \cite{krevor2015capillary}. 

It is important to consider the possible limitations of this model for such realistic scenarios. First and foremost, it must be noted that the anisotropic permeability values $k_H$, $k_V$, are upscaled quantities which attempt to capture the macroscopic effect of small-scale heterogeneities on the flow. These upscaled quantities are only a good approximation when the vertical length scale of the flow $H$ is much larger than the heterogeneity length scale (e.g. the width of sedimentary layers) or when the permeable interval is inherently anisotropic due to compaction effects. If this is not the case, more complex flow models are required to treat the spread of fluid beneath and through successive layers \cite{neufeld2009modelling,huppert2014fluid,hewitt2022evolution}. 

Another consideration for the case of CO$_2$ sequestration is the effect of trapped saturation on the dynamics and spreading of the current. As the current moves, a fraction of its mass is lost to residual trapping due to small scale capillary forces \cite{krevor2015capillary} and dissolution within the surrounding brine \cite{macminn2011co}. According to some trapping models \cite{golding2017two} this can arrest the spread of the current altogether. Likewise, the spreading could also be arrested by lateral heterogeneities in the capillary pressure pinning the nose of the gravity current. 

Whilst the conclusion of this study is that anisotropy reduces the swept volume of the gravity current, this doesn't take into account the enhanced trapping potential due to changes in the capillary pressure across heterogeneities, also known as capillary heterogeneity trapping \cite{harris2021impact}. Specifically, during the imbibition cycle small-scale capillary forces induce a build-up of saturation beneath each sedimentary layer that can account for as much as 14$\%$ of the overall trapped saturation. Hence, the optimum anisotropy will no doubt strike a balance between the trapping associated with these heterogeneities and the reduction in swept volume that they induce. 

It is also worth mentioning the flow of the ambient fluid, which we have so far ignored for this study.  
In particular, wherever there are sharp changes in the interface shape, this may be associated with non-negligible displacement of the local ambient fluid, such that the dynamic boundary condition \eqref{dyn} cannot be applied. Likewise, in the case where there is a viscosity contrast between the released and ambient fluids, this can modify the shape of the current \cite{pegler2014fluid} and cause fingering instabilities \cite{saffman1958penetration}. In both of these cases a full numerical model would be necessary to resolve such flow details. 
In the case of carbon sequestration, CO$_2$ is typically 20-30 times less viscous than brine. As shown by \cite{pegler2014fluid}, the viscosity contrast causes an enhanced spreading of the CO$_2$ in the shape of a thin finger along the cap rock. Hence, this viscosity contrast distorts the spreading of CO$_2$ in a similar manner to anisotropy, as studied here. Therefore, the expected effect of a viscosity contrast in anisotropic media is an extremely pronounced finger-like intrusion of CO$_2$.

%A future study could consider a dynamic injection rate $Q(t)$ for a given heterogeneous or anisotropic aquifer. In such cases it would be interesting to optimise $Q(t)$ so as to maximise trapping, either by way of increasing the swept volume, or by exploiting capillary heterogeneity trapping associated with sedimentary layers. 

\dataccess The finite difference code used in this study can be found in the Supplemental Materials or on the personal website of the author: \url{https://people.maths.ox.ac.uk/benham/openacces.zip}
%\aucontribute G.P.B. is the sole author. 
\competing The author reports no competing interests.
\funding There is no funding to report. 
\ack The author wishes to thank the anonymous peer reviewers for time taken to review the manuscript.

\appendix

\section{Additional plots}
\label{sec_appa}

\begin{figure}
\centering
\begin{tikzpicture}[scale=1.1]
\node at (0,0) {\includegraphics[width=0.32\textwidth]{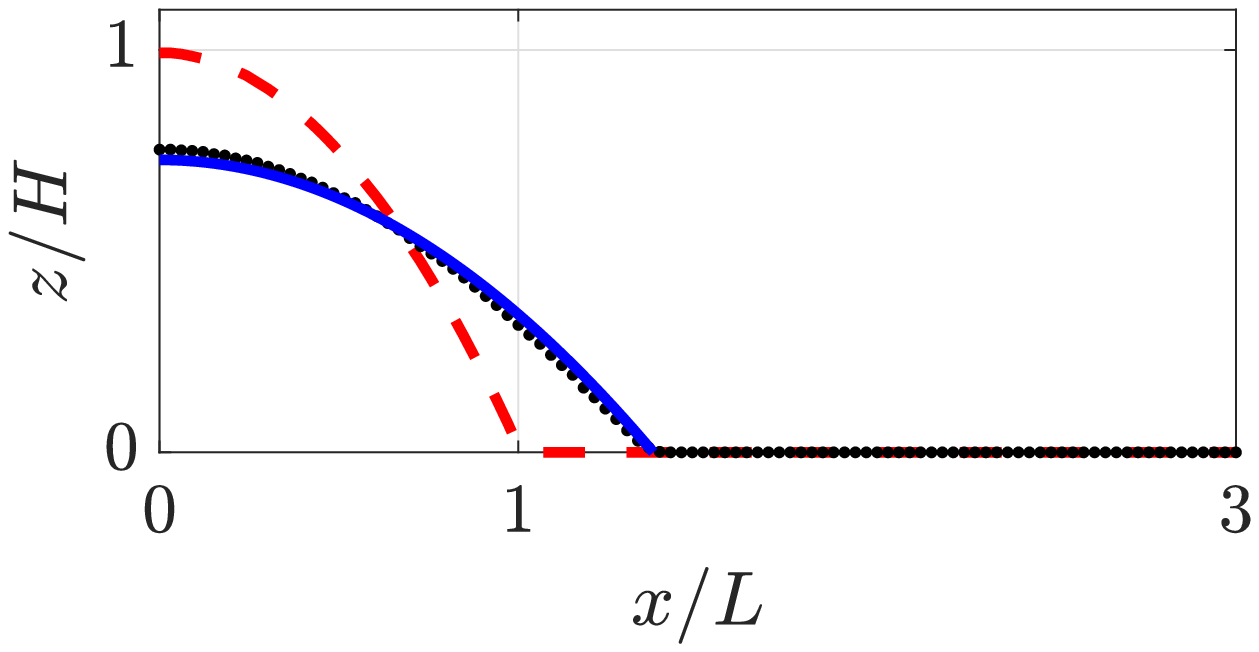}};
\node at (4.1,0) {\includegraphics[width=0.32\textwidth]{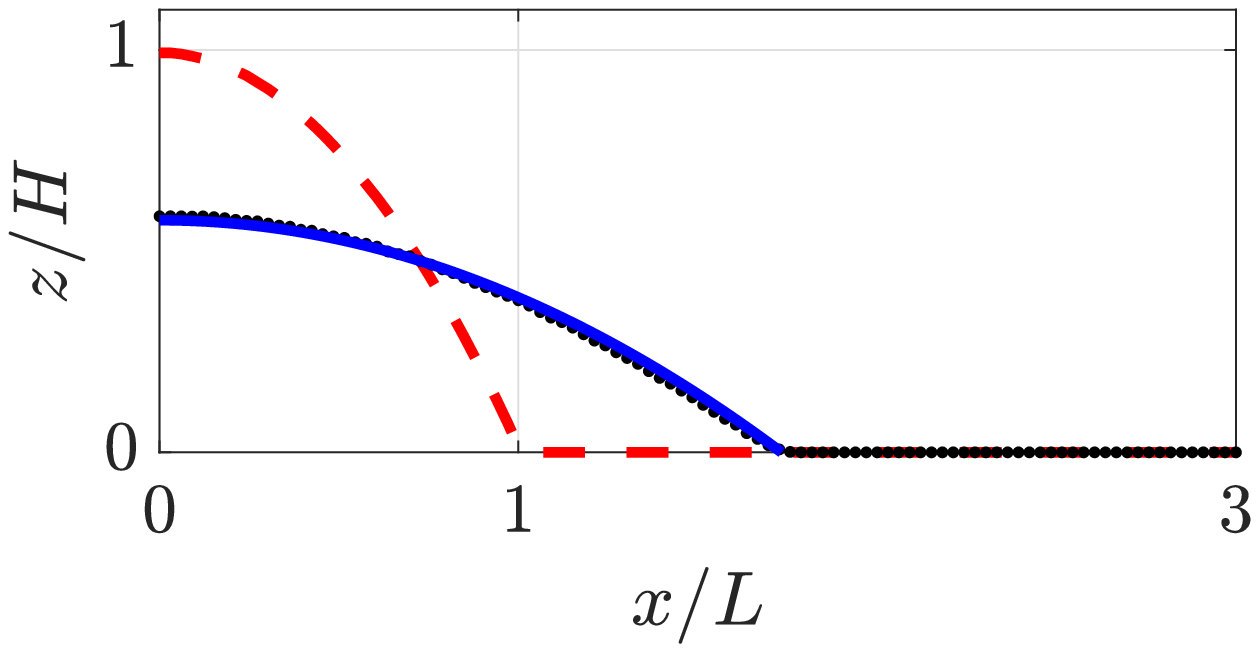}};
\node at (8.25,0) {\includegraphics[width=0.32\textwidth]{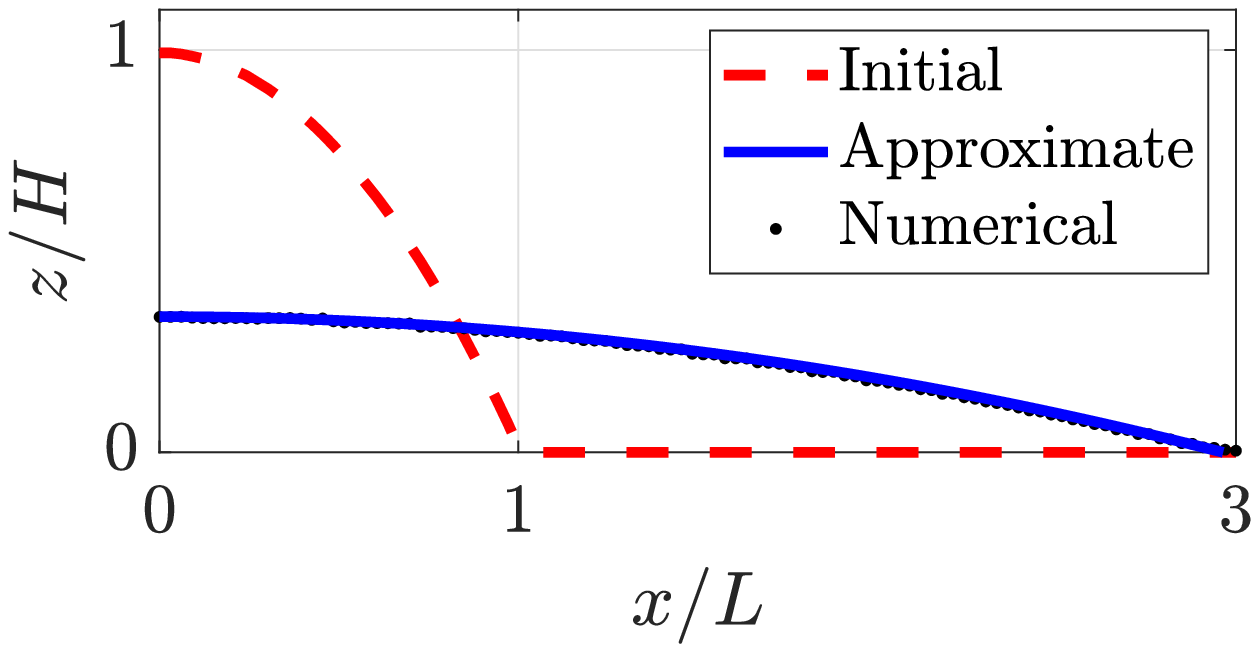}};
\node at (0,-2.5) {\includegraphics[width=0.32\textwidth]{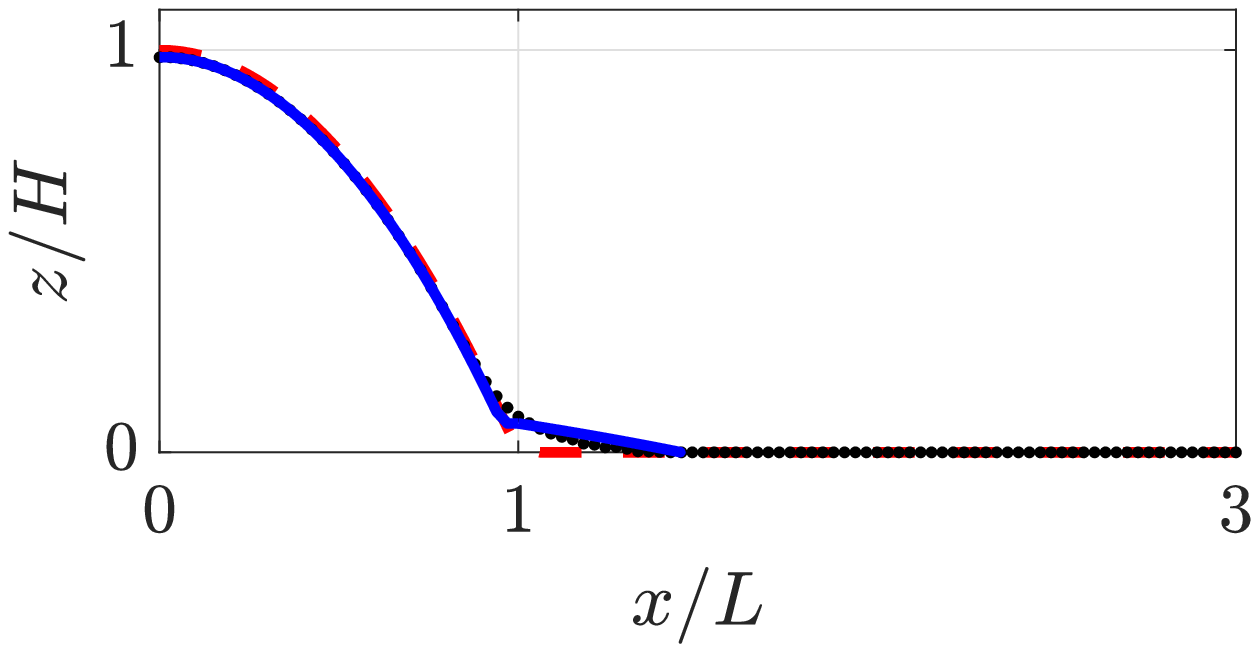}};
\node at (4.1,-2.5) {\includegraphics[width=0.32\textwidth]{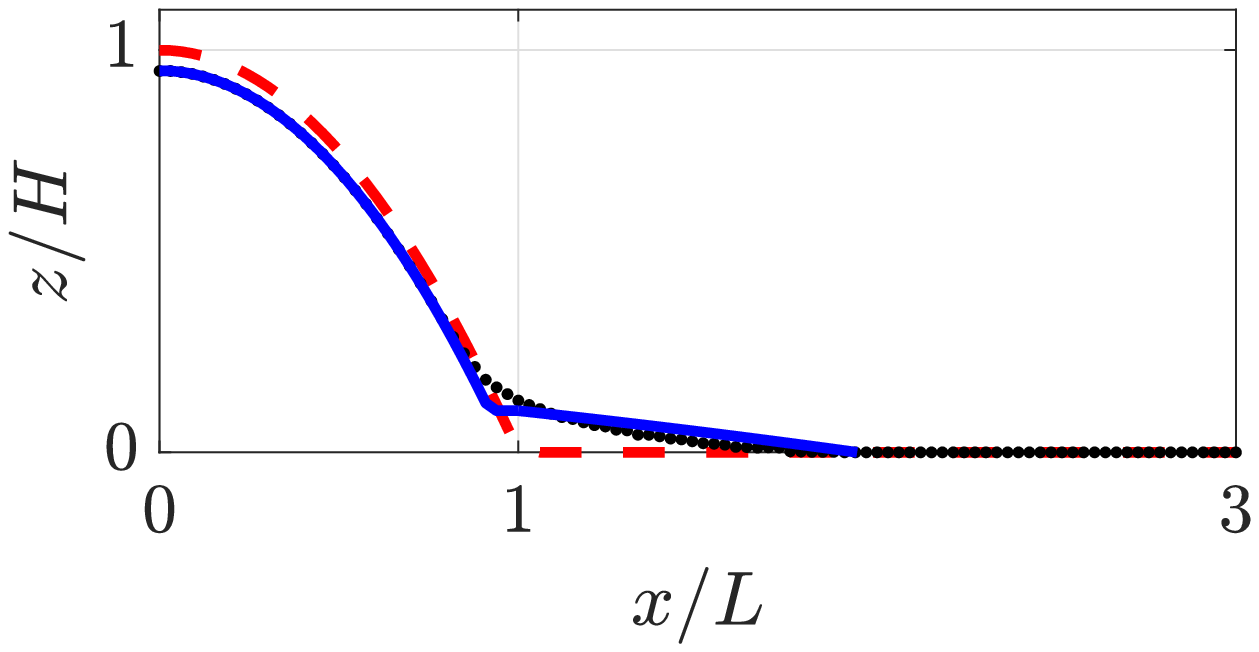}};
\node at (8.25,-2.5) {\includegraphics[width=0.32\textwidth]{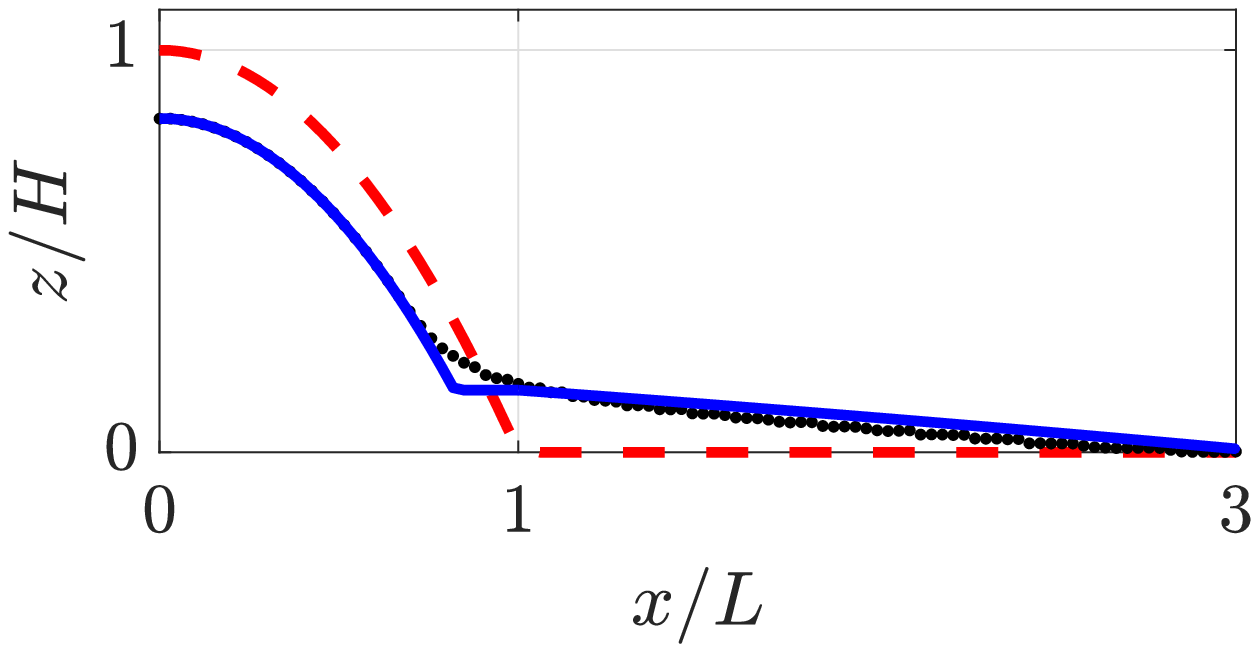}};
\node at (-1.6,1.4) {(a)};
\node at (2.7,1.4) {(b)};
\node at (6.8,1.4) {(c)};
\node at (-1.6,-1.1) {(d)};
\node at (2.7,-1.1) {(e)};
\node at (6.8,-1.1) {(f)};
\end{tikzpicture}
\caption{Evolution of a gravity current with initial shape given by \eqref{initsim} in the case of an isotropic medium $\epsilon=1$ (a,b,c) and an anisotropic medium $\epsilon=0.01$ (d,e,f). Dimensionless times are given by $u_b t/L\phi=$ 0.87, 1.74, 8.68 in (a,b,c) and 1.70, 5.09, 17.0 in (d,e,f). 
\label{num_app}}
\end{figure}

In this section we present some additional plots comparing the numerical solution from Section \ref{sec_num} with various approximate and analytical solutions. 
Whilst in the previous sections a rectangular shape was chosen for the initial released shape of dense fluid, here we consider a curved profile with initial shape $h_0(x)$ given by \eqref{initsim}. In the case of an isotropic medium $\epsilon=1$, this results in immediate self-similar behaviour, as described in Section \ref{sec_trans}. We use the analytical self-similar solution \eqref{analf} as a means of verifying our numerical method. In figure \ref{num_app}a,b,c, profiles of the gravity current are shown at three different times. Overall, very good agreement is found, indicating the soundness of the numerical method.

Unfortunately, no benchmark analytical solution exists for anisotropic porous media, but we can nevertheless compare against the approximate solution derived in this study. In figure \ref{num_app}d,e,f plots are shown for the same curved initial shape \eqref{initsim} released in an anisotropic medium with $\epsilon=0.01$. It is straightforward to extend the approximate solution derived earlier to this initial shape. As such, the bulk fluid region I initially evolves according to
\beq
h=h_0(x)- \frac{\epsilon u_b t}{\phi}.\label{lintime2}
\eeq
Meanwhile, region II initially evolves according to the self-similar dynamics \eqref{earlysim} which correspond with a finger-like region fed by a constant input flux  $\epsilon u_b L$ from region I (see Section \ref{sec_early}).  These two solutions are then simply joined together to make the plots in figure \ref{num_app}d,e,f. Overall, good agreement is achieved indicating that our model can be extended to this and other such similar initial shapes.

\section{Further details on similarity solutions}
\label{sec_appb}

In this section, we summarise the equations that define the different similarity solutions used in the main text. 
Let's start with the two-dimensional equations for region II at early times. In this case the system of equations deriving from \eqref{massconIII}-\eqref{hbc2} after applying coordinate transformations \eqref{simvars1}-\eqref{earlysim} is 
\begin{align}
\frac{1}{3}f-\frac{2}{3}\eta \frac{\mathrm{d}f}{\mathrm{d}\eta} & = \frac{\mathrm{d}}{\mathrm{d}\eta}\left[ f \frac{\mathrm{d}f} {\mathrm{d}\eta}\right],\\
-f\frac{\mathrm{d}f}{\mathrm{d}\eta}&=1:\quad \eta=0,\\
f&=0:\quad \eta=\eta_N,\\
-f\frac{\mathrm{d}f}{\mathrm{d}\eta}&=0:\quad \eta=\eta_N.
\end{align}
These can be solved numerically for the shape function $f(\eta)$ and prefactor $\eta_N=1.482$.

Next we summarise the two-dimensional equations at late times, once the gravity current has transitioned to a single slumping region. In this case the system of equations \eqref{massconIII},\eqref{hbc1},\eqref{symbc},\eqref{intconst}, after applying coordinate transformations \eqref{finitescle0}-\eqref{finitescle}, becomes
\begin{align}
-\frac{1}{3}f-\frac{1}{3}\eta \frac{\mathrm{d}f}{\mathrm{d}\eta} & = \frac{\mathrm{d}}{\mathrm{d}\eta}\left[ f \frac{\mathrm{d}f} {\mathrm{d}\eta}\right],\\
f&=0:\quad \eta=\eta_N,\\
\frac{\mathrm{d}f}{\mathrm{d}\eta}&=0:\quad \eta=0,\\
\int_0^{\eta_N} f \,\mathrm{d}\eta & = 1.
\end{align}
These can be solved analytically to give \eqref{analf} and $\eta_N=3^{2/3}$.

In the radially symmetric case, the governing equations for region II at early times are \eqref{massconrad}-\eqref{hbcrad2}. Hence, after applying the coordinate transformations \eqref{simvarsrad1}-\eqref{earlysimrad}, we get the system of equations
\begin{align}
-\frac{1}{2}\eta \frac{\mathrm{d}f}{\mathrm{d}\eta} & = \lb \frac{1}{1+\eta \tau^{1/2}}\rb\frac{\mathrm{d}}{\mathrm{d}\eta}\left[ (1+\eta \tau^{1/2}) f \frac{\mathrm{d}f} {\mathrm{d}\eta}\right],\\
-2\pi\lb\frac{1+\eta \tau^{1/2}}{\tau^{1/2}}\rb f\frac{\mathrm{d}f}{\mathrm{d}\eta}&=1:\quad \eta=0,\\
f&=0:\quad \eta=\eta_N,\\
-2\pi\lb\frac{1+\eta \tau^{1/2}}{\tau^{1/2}}\rb f\frac{\mathrm{d}f}{\mathrm{d}\eta}&=0:\quad \eta=\eta_N.
\end{align}
After sufficiently long times $1\ll \tau\ll \tau^*$ (where $\tau^*$ is the dimensionless transition time), the time-dependence is removed from the above system. In other words, we consider when enough time has passed that the finger has grown far from the initial shape, but not so long for transition to occur. In this case, the system of equations simplifies to
\begin{align}
-\frac{1}{2}\eta \frac{\mathrm{d}f}{\mathrm{d}\eta} & = \frac{1}{\eta}\frac{\mathrm{d}}{\mathrm{d}\eta}\left[ \eta f \frac{\mathrm{d}f} {\mathrm{d}\eta}\right],\\
-2\pi\eta f\frac{\mathrm{d}f}{\mathrm{d}\eta}&=1:\quad \eta=0,\\
f&=0:\quad \eta=\eta_N,\\
-2\pi\eta f\frac{\mathrm{d}f}{\mathrm{d}\eta}&=0:\quad \eta=\eta_N.
\end{align}
These can be solved numerically for the shape function $f(\eta)$ and prefactor $\eta_N=1.155$.

Finally we summarise the radially symmetric equations at late times, once the gravity current has transitioned to a single slumping region. In this case the system of equations \eqref{massconrad},\eqref{hbcrad1},\eqref{symbcrad},\eqref{intconstrad}, after applying coordinate transformations \eqref{finitesclerad0}-\eqref{finitesclerad}, becomes 
\begin{align}
-\frac{1}{2}f-\frac{1}{4}\eta \frac{\mathrm{d}f}{\mathrm{d}\eta} & =  \frac{1}{\eta}\frac{\mathrm{d}}{\mathrm{d}\eta}\left[ \eta f \frac{\mathrm{d}f} {\mathrm{d}\eta}\right],\\
f&=0:\quad \eta=\eta_N,\\
\frac{\mathrm{d}f}{\mathrm{d}\eta}&=0:\quad \eta=0,\\
2\int_0^{\eta_N} \eta f \, \mathrm{d}\eta & = 1.
\end{align}
These can be solved analytically to give \eqref{analfrad} and $\eta_N=2$.

\enlargethispage{20pt}

%\ack{Insert acknowledgment text here.}

\vskip2pc

\bibliographystyle{RS}
\bibliography{bibfile}

\end{document}